\documentclass[aps, prd, twocolumn, superscriptaddress, nofootinbib, showpacs,preprintnumbers]{revtex4}
\usepackage{graphicx}
\usepackage{amsmath}
\usepackage{mathrsfs}
\usepackage{ulem} 
\usepackage{color}

\def\fsl#1{\setbox0=\hbox{$#1$}                 
   \dimen0=\wd0                                 
   \setbox1=\hbox{/} \dimen1=\wd1               
   \ifdim\dimen0>\dimen1                        
      \rlap{\hbox to \dimen0{\hfil/\hfil}}      
      #1                                        
   \else                                        
      \rlap{\hbox to \dimen1{\hfil$#1$\hfil}}   
      /                                         
   \fi}                                         %
\newcommand{\tr}{\mbox{tr}}

\newcommand{\VEV}[1]{\langle #1 \rangle}
\newcommand{\gtrsim}{\mathop{>}\limits_{\displaystyle{\sim}}}

\newcommand{\unit}{\mathbf{1}}
\makeatletter
\@addtoreset{equation}{section}
\makeatother

\begin{document}
\preprint{KEK-TH-1844}
\title{
Unitarity sum rules, three site moose model,
and the ATLAS 2 TeV diboson anomalies
}  
\date{\today}

\author{Tomohiro Abe}
\email[]{abetomo@post.kek.jp}
\affiliation{
Institute of Particle and Nuclear Studies,
High Energy Accelerator Research Organization (KEK),
Tsukuba 305-0801, Japan
}

\author{Ryo Nagai}
\email[]{nagai@eken.phys.nagoya-u.ac.jp}
\affiliation{
  Department of Physics, Nagoya University, Nagoya 464-8602, Japan
}

\author{Shohei Okawa}
\email[]{okawa@eken.phys.nagoya-u.ac.jp}
\affiliation{
  Department of Physics, Nagoya University, Nagoya 464-8602, Japan
}

\author{Masaharu Tanabashi}
\email[]{tanabash@eken.phys.nagoya-u.ac.jp}
\affiliation{
  Department of Physics, Nagoya University, Nagoya 464-8602, Japan
}
\affiliation{
  Kobayashi-Maskawa Institute for the Origin of Particles and the Universe, 
  Nagoya University, Nagoya 464-8602, Japan
}

\begin{abstract}
We investigate $W'$ interpretations for the ATLAS 2 TeV diboson 
anomalies.
The roles of the unitarity sum rules, which 
ensure the perturbativity of the longitudinal vector boson scattering 
amplitudes, are emphasized.
We find the unitarity sum rules and the custodial symmetry 
are powerful enough to predict various nontrivial 
relations among
$WWZ'$, $WZW'$, $WWh$, $WW'h$ and $ZZ'h$ coupling strengths 
in a model independent manner.
We also perform surveys in the general parameter space of 
$W'$ models and find 
the ATLAS 2 TeV diboson anomalies may be interpreted as
a $W'$ particle of the three site moose model, i.e., a
Kaluza-Klein like particle in a deconstructed extra dimension model.
It is also shown that the non standard-model-like Higgs boson is favored by the
present data to interpret the ATLAS diboson anomalies as the
consequences of the $W'$ and $Z'$ bosons.
\end{abstract}

\maketitle

\section{Introduction}
\label{sec:introduction}

Recently, the ATLAS Collaboration of the LHC experiment reported
anomalies in their search for high-mass diboson ($WZ$, $WW$ or $ZZ$) 
resonances with 
boson-tagged jets 
at the diboson invariant mass 2 TeV~\cite{Aad:2015owa}.
$W$ and $Z$ bosons resulting from high-mass resonance are highly 
boosted, so each boson's hadronic decay products are reconstructed
as a single fat jet $J$ in this search.  
The reported 
local significances of anomalies are $3.4$ $\sigma$, $2.6$ $\sigma$,
and $2.9$ $\sigma$ for $WZ\to JJ$, $WW \to JJ$, and 
$ZZ\to JJ$ channels, respectively.
If we explain the ATLAS $WZ\to JJ$ anomaly in the $W'$ model, we 
need to introduce a narrow high-mass $W'$ boson with $M_{W'}=2$ TeV, 
$\Gamma_{W'} < 100$ GeV, and 
\begin{equation}
 \sigma(pp\to W'; \sqrt{s}=8\mbox{ TeV}) B_{W'}(WZ)\simeq 14 \mbox{ fb}. 
\end{equation}
Note here the above cross section is the best fit value. 
A $W'$ particle with a little bit smaller production cross section
may also be consistent with the ATLAS diboson anomaly.

The CMS Collaboration also reported their results of search for
high-mass diboson resonance in the same decay 
channel~\cite{Khachatryan:2014hpa}. 
Although the CMS Collaboration reported a small excess 
around 1.8 TeV, at the resonance mass of 2 TeV, CMS only gives the 
upper limit on $\sigma B$,
$\sigma(pp\to W'; \sqrt{s}=8\mbox{ TeV}) B_{W'}(WZ) < 13$ fb.
The ATLAS Collaboration also reported their search for high-mass
diboson resonance $X$ in the $WV \to \ell\nu q\bar{q}$ decay 
channel~\cite{Aad:2015ufa}.
Here $V$ stands for $W$ or $Z$.
The ATLAS limit on $\sigma B$ in this 
decay channel is about 
\begin{equation}
  \sum_{X} \sigma(pp\to X; \sqrt{s}=8\mbox{ TeV}) B_{X}(W V) 
  < 10 \mbox{ fb}, 
\label{eq:tension}
\end{equation}
for 2 TeV narrow resonance $X$.
The limit (\ref{eq:tension}) causes a tension with the 
best fit value  of the ATLAS 
diboson anomaly $\sigma(pp\to W') B_{W'}(WZ)\simeq 14$ fb.
We use the value $\sigma B \simeq 10$ fb 
as a reference for our interpretation of the ATLAS diboson anomaly 
in this paper, though only 70\% of the 
ATLAS diboson excess can be explained with this cross section.

The CMS Collaboration recently reported their search limit on 
the $W'$ production in the $W' \to Wh \to JJ$ decay 
channel in Ref.\cite{Khachatryan:2015bma}.
Although the CMS Collaboration found a small excess of event 
numbers around $M_{W'}=1.8$TeV, they found rather severe upper 
limit for $M_{W'}=2$ TeV resonance, i.e., about 
$\sigma(pp \to W'; \sqrt{s}=8\mbox{ TeV}) B_{W'}(Wh) < 7$ fb.
As we will see later, this upper limit causes a severer tension
with the ATLAS diboson anomaly.
In Ref.\cite{kn:CMS_higgs2}, the CMS Collaboration 
reported their search result for high-mass resonance $W'$ 
in the decay channel $W' \to Wh \to \ell\nu b\bar{b}$.  
Again, they found an excess at $M_{W'} \simeq 1.8$ TeV.

After the ATLAS Collaboration reported the 2 TeV diboson 
anomalies~\cite{Aad:2015owa},
many studies of possible theoretical interpretations have 
appeared in the 
market~\cite{Fukano:2015hga,Hisano:2015gna,Franzosi:2015zra,Cheung:2015nha,Dobrescu:2015qna,Aguilar-Saavedra:2015rna,Alves:2015mua,Gao:2015irw,Thamm:2015csa,Brehmer:2015cia,Cao:2015lia,Cacciapaglia:2015eea}.
One of the biggest questions raised in these interpretations 
is whether the ATLAS diboson anomaly 
at $M=2$ TeV is related with the CMS excesses at $M=1.8$ TeV or not.
Given the situation where the jet mass resolutions
of ATLAS and CMS detectors are much better than 100 GeV, 
it seems unlikely that the CMS 1.8 TeV excesses are directly related 
with the ATLAS 2 TeV anomalies, however.
If this is not so, the next theoretical challenge is 
to make viable models of $W'$ explaining the ATLAS diboson 
anomalies 
without 
causing conflicts with the CMS upper limit on the 
$W' \to Wh$ decay channel, 
$\sigma(pp \to W') B_{W'}(Wh) < 7$ fb at $M_{W'}=2$ TeV\@.
This is a tough challenge, however, if we take the CMS 
7 fb upper limit seriously.
The Higgs boson is the $SU(2)_W$ partner of the would-be 
Nambu-Goldstone  boson (NGB) in the Standard Model (SM).
The equivalence theorem of the longitudinal $W$ boson and the
eaten would-be NGB amplitudes then suggests us a relation
\begin{equation}
  \Gamma_{W'}(WZ) = \Gamma_{W'}(Wh)
\label{eq:equivalence}
\end{equation}
for sufficiently heavy $W'$.
See, e.g, Ref.~\cite{Pappadopulo:2014qza} for a typical $W'$ model
satisfying this relation (\ref{eq:equivalence}).
The relation (\ref{eq:equivalence}) implies that the 
$W'$ branching fraction to the $Wh$ mode is identical to
the $W'$ branching fraction to $WZ$, i.e., $B_{W'}(WZ)=B_{W'}(Wh)$.
The CMS 7 fb upper limit on the $Wh$ channel therefore gives an
upper limit on $\sigma(pp \to W') B(WZ)<7$ fb.
Less than only a half of the ATLAS diboson anomaly excess $\sim 14$ fb 
can be explained!

Although this tension may be explained by a statistical fluctuation
at the present stage,
it is tempting to consider scenarios free from the relation 
(\ref{eq:equivalence}).
In this paper, we point out that the relation (\ref{eq:equivalence}) 
is true only if the couplings of the 125 GeV Higgs boson 
with $WW$ and $ZZ$ are same as the SM predictions.
We investigate the relation (\ref{eq:equivalence})
from the viewpoint of the perturbative unitarity of the
longitudinal $W$ and $W'$ boson scattering amplitudes.
We derive a set of unitarity sum rules among coupling
strengths of $W'$ and $W$ bosons, which should be satisfied
in any perturbative model of $W'$.
We then obtain a relation among the $WZW'$ coupling, 
the $WWh$ coupling, and the $WhW'$ coupling by using these
unitarity sum rules.
We find that, if the 125 GeV Higgs is a non-SM Higgs boson,
the relation (\ref{eq:equivalence}) should be modified as
\begin{equation}
  \kappa_V^2 \Gamma_{W'}(WZ) = \Gamma_{W'}(Wh) .
\label{eq:equivalence2}
\end{equation}
Here $\kappa_V$ is defined as $\kappa_V = g_{WWh}/g_{WWh}^{\rm SM}$,
with $g_{WWh}$ and $g_{WWh}^{\rm SM}=g_W M_W$ being
the $WWh$ coupling strength and its corresponding SM value, 
respectively.
The ATLAS 2 TeV diboson anomalies may be consistent with 
the CMS limits on the $Wh$ decay channel, 
if we consider a model with $\kappa_V < 1$.~\footnote{
Here we assume existence of mechanism to adjust the Higgs 
measurement signal strengths with $\kappa_V < 1$.
}

Inspired by the unitarity sum rules and the custodial 
$SU(2)$ symmetry arguments, we then 
introduce a parametrization of $W'$ and $Z'$ couplings,
and survey the parameter space to find phenomenologically
viable models consistent with the existing limits on the
$W'$ and $Z'$ particles.
We then show that the three site moose model~\cite{Abe:2012fb}, 
a linear sigma model generalization of the three site Higgsless 
model~\cite{Chivukula:2006cg}, can explain the parameter space
consistent both with the ATLAS anomalies and with 
the existing limits on $W'$ and $Z'$.
Note that the three site Higgsless model is a 
deconstruction\cite{ArkaniHamed:2001ca,Hill:2000mu}
version of the extra dimension Higgsless model\cite{Csaki:2003dt}.
The gauge symmetry breaking structure of the three 
site moose model thus resembles the structure of extra dimension
models containing bulk weak gauge fields.
The $W'$ and $Z'$ bosons in the three site moose model can 
therefore be regarded as the Kaluza-Klein (KK) modes of the weak gauge 
bosons.
We also note that, 
as emphasized in Ref.~\cite{Abe:2012fb}, the three site moose 
model implements a mechanism to adjust the Higgs signal strengths
even with $\kappa_V < 1$.

This paper is organized as follows:
In Sec.~\ref{sec:unitarity}, we derive a set of unitarity sum rules
in a class of models with the custodial symmetry 
including arbitrary number of 
custodial $SU(2)$ triplet vector bosons ($W, W', W'', \cdots$) 
and neutral Higgs bosons ($h_1, h_2, \cdots$).
We obtain a relationship between the $WW'h$ coupling and 
the $WWh$ coupling by using the unitarity sum rules.
We propose a parametrization for the $W'$ and $Z'$ couplings 
in a manner consistent with
the perturbativity and the custodial 
symmetry in Sec.~\ref{sec:parametrization}.
Surveys in the parameter space of $W'$ and $Z'$ models 
are presented 
in Sec.~\ref{sec:fit}.
Sec.~\ref{sec:threesite} is devoted to the three site moose 
model.  
Summary and outlooks are presented in Sec.~\ref{sec:summary}.

\section{Unitarity sum rules}
\label{sec:unitarity}

In order to keep the perturbative unitarity in the
longitudinal vector boson scattering amplitudes,
in any perturbative model,
self-interaction coupling strengths 
of massive vector bosons
need to satisfy a set of unitarity sum 
rules~\cite{Cornwall:1973tb,Cornwall:1974km,Llewellyn Smith:1973ey,Gunion:1990kf}. 
Examples of such unitarity sum rules 
for a model including
a tower of massive vector bosons ($W, W', W'',\cdots$)
are presented 
in Ref.~\cite{SekharChivukula:2008mj}
in the context of the deconstructed Higgsless theory.
Unitarity sum rules in a model with $W, W'$ and neutral Higgs
bosons ($h_1, h_2, \cdots$) are discussed in 
Ref.~\cite{Abe:2012fb}.
See also Refs.~\cite{Foadi:2008xj,Hernandez:2010iu,Bellazzini:2012tv,Englert:2015oga}.
In this section, we further generalize the unitarity sum rules
of Ref.~\cite{Abe:2012fb} to obtain a relationship between 
$WWh$ and $WW'h$ couplings.

\subsection{General Sum Rules}
\label{sec:general-sum-rules}

For simplicity, in this section, 
we consider the custodial $SU(2)$ symmetry limit.
Effects of the custodial symmetry violation arising from the
weak hypercharge gauge coupling will be discussed later.
The model we consider contains $N_V$ custodial $SU(2)$ triplet vector
bosons ($W^a_{i\mu}$, $a=1,2,3$, $i=0, 1, \cdots, N_V-1$) and
$N_h$ singlet Higgs bosons ($h_i$, $i=1,2, \cdots, N_h$).
In order to cancel the $E^4$-behavior of the longitudinal
$W_i W_j \to W_k W_\ell$ scattering amplitudes,
quartic vector boson coupling strengths
$g_{W_i W_j W_k W_\ell}$ need to satisfy
\begin{equation}
  g_{W_i W_j W_k W_\ell} = \sum_m g_{W_i W_j W_m} g_{W_k W_\ell W_m},
\label{eq:e4-sr1}
\end{equation}
with $g_{W_i W_j W_k W_\ell}$ being symmetric under the exchange of
the indices $i$, $j$, $k$, $\ell$ (Bose symmetry).
The triple vector boson couplings also satisfy
the Bose symmetry.
Here we specified the quartic and triple vector couplings by using
notations similar to Ref.~\cite{Abe:2012fb}.
Using (\ref{eq:e4-sr1}) and the Bose symmetry of $g_{W_i W_j W_k W_\ell}$, 
it is easy to see
\begin{eqnarray}
 \sum_m g_{W_i W_j W_m} g_{W_k W_\ell W_m}
 &=& \sum_m g_{W_i W_k W_m} g_{W_j W_\ell W_m}
 \nonumber\\
 &=& \sum_m g_{W_i W_\ell W_m} g_{W_k W_j W_m}.
 \nonumber\\
 & &
\label{eq:e4-sr2}
\end{eqnarray}

Requiring the cancellation of the $E^2$-behavior, we also 
obtain a sum rule which relates
the Higgs coupling $g_{W_i W_j h_k}$ with
the vector boson self-interaction couplings,
\begin{eqnarray}
\lefteqn{
  \sum_m g_{W_i W_j h_m} g_{W_k W_\ell h_m}
} \nonumber\\
  &=& (M_{W_i}^2+M_{W_j}^2+M_{W_k}^2+M_{W_\ell}^2) g_{W_i W_j W_k W_\ell}
  \nonumber\\
  & & - \sum_m M_{W_m}^2 g_{W_i W_j W_m} g_{W_k W_\ell W_m}
  \nonumber\\
  & &
      - \sum_m M_{W_m}^2 g_{W_i W_k W_m} g_{W_j W_\ell W_m}
  \nonumber\\
  & &
      - \sum_m M_{W_m}^2 g_{W_i W_\ell W_m} g_{W_k W_j W_m} .
\label{eq:e2-sr1}
\end{eqnarray}
Again we used notations similar to Ref.~\cite{Abe:2012fb}.
We also obtain
\begin{eqnarray}
\lefteqn{
\sum_m \dfrac{(M_{W_i}^2 - M_{W_j}^2) (M_{W_k}^2 - M_{W_\ell}^2)}{M_{W_m}^2} 
       g_{W_i W_j W_m} g_{W_k W_\ell W_m}
} \nonumber\\
  &=& \sum_m M_{W_m}^2 \biggl(
        -g_{W_i W_k W_m} g_{W_j W_\ell W_m} 
  \nonumber\\
  & & \qquad\qquad\qquad\qquad
        +g_{W_i W_\ell W_m} g_{W_k W_j W_m}
      \biggr).
\label{eq:e2-sr2}
\end{eqnarray}
Note here that the RHS of (\ref{eq:e2-sr1})  is symmetric under the
exchange of the indices $i$, $j$, $k$, $\ell$.  
We therefore obtain
\begin{eqnarray}
 \sum_m g_{W_i W_j h_m} g_{W_k W_\ell h_m}
 &=& \sum_m g_{W_i W_k h_m} g_{W_j W_\ell h_m}
 \nonumber\\
 &=& \sum_m g_{W_i W_\ell h_m} g_{W_k W_j h_m}.
 \nonumber\\
 & &
\label{eq:e2-sr3}
\end{eqnarray}

\subsection{Properties of $W'$}
\label{sec:properties-w}

We are now ready to discuss applications of the unitarity sum rules
(\ref{eq:e4-sr1}), (\ref{eq:e4-sr2}), (\ref{eq:e2-sr1}), 
(\ref{eq:e2-sr2}) and (\ref{eq:e2-sr3}).

We first consider the sum rule which ensures the cancellation of the
$E^4$-behavior in the $WW \to WW$ amplitude.
Eq.(\ref{eq:e4-sr1}) reads
\begin{equation}
  g_{WWWW} = g_{WWW}^2 + g_{WWW'}^2 .
\label{eq:e4-wwww}
\end{equation}
We assume here that the sum rule is saturated only by $W$ ($=W_0$) 
and $W'$ ($=W_1$).
Effects of possibly existing heavier resonances $W''$ ($=W_2$), $\cdots$,
are assumed to be negligible.

We should note that the sum rule (\ref{eq:e4-wwww}) is incomplete below
the energy scale $E \lesssim M_{W'}$,
where the $W'$ exchange term $g_{WWW'}^2$ does not affect the $WW \to WW$ 
amplitude.
The longitudinal polarization of $W$ gives a factor $E/M_W$ in
the amplitude for $E \gg M_W$.
For the energy scale $M_{W'} \gtrsim E \gg M_W$, the
$WW \to WW$ amplitude therefore behaves as
\begin{equation}
  \left( g_{WWWW} - g_{WWW}^2 \right) \dfrac{E^4}{M_W^4}.
\end{equation}
Requiring the amplitude is still in a perturbative 
regime~\cite{Lee:1977eg}
at $E=M_{W'}$, we obtain a condition
\begin{equation}
  \left( g_{WWWW} - g_{WWW}^2 \right) \dfrac{M_{W'}^4}{M_W^4} \lesssim
  32\pi .
\end{equation}
It is now easy to see that the $g_{WWW'}$ coupling need to satisfy
\begin{equation}
  g_{WWW'}^2 \dfrac{M_{W'}^4}{M_W^4} \lesssim 32\pi.
\label{eq:wwww-perturbativity}
\end{equation}
Parametrizing the $g_{WWW'}$ coupling
\begin{equation}
  g_{WWW'} = \xi_V g_{WWW} \dfrac{M_W^2}{M_{W'}^2},
\label{eq:wwwp-parametrization}
\end{equation}
the perturbativity condition (\ref{eq:wwww-perturbativity}) 
can be expressed as
\begin{equation}
  | \xi_V | \lesssim 15.
\label{eq:wwww-perturbativity2}
\end{equation}
Here we used $g_{WWW} \simeq 0.65$.
In typical analyses of collider phenomenologies of $W'$,
the parameter $\xi_V$ is chosen to be 
$\xi_V\simeq 1$.
Although this choice clearly satisfies the perturbativity
condition (\ref{eq:wwww-perturbativity2}),
it is also possible to construct models with larger value of
$\xi_V$, e.g, $\xi_V \sim 5$, still keeping the perturbativity
condition (\ref{eq:wwww-perturbativity}).

We next consider the sum rule which ensures the cancellation
of the $E^4$-behavior in the $WW \to W'W'$ amplitude,
Eq.(\ref{eq:e4-sr2}),
\begin{equation}
  g_{WW'W}^2 + g_{WW'W'}^2 = g_{WWW} g_{W'W'W} + g_{WWW'} g_{W'W'W'}.
\label{eq:wwwpwp-unitarity}
\end{equation}
Eq.(\ref{eq:wwwpwp-unitarity}) can be regarded as a quadratic
equation of $g_{WW'W'}$,
\begin{eqnarray}
  0 &=& g_{WW'W'}^2 - g_{WWW} g_{WW'W'} 
    \nonumber\\
    & & + g_{WWW'}^2 - g_{W'W'W'} g_{WWW'}.
\label{eq:wwwpwp-quad}
\end{eqnarray}
Plugging (\ref{eq:wwwp-parametrization}) into (\ref{eq:wwwpwp-quad}),
we obtain
\begin{eqnarray}
  0 &=& g_{WW'W'}^2 - g_{WWW} g_{WW'W'}
    \nonumber\\
    & & + \xi_V^2 g_{WWW}^2 \dfrac{M_W^4}{M_{W'}^4} 
    \nonumber\\
    & &
        - \xi_V g_{W'W'W'} g_{WWW} \dfrac{M_W^2}{M_{W'}^2}  .
\label{eq:wwwpwp-quad2}
\end{eqnarray}
Solving the quadratic equation (\ref{eq:wwwpwp-quad2}) in the
$M_{W'}^2 \gg M_W^2$ limit,
we obtain two solutions
\begin{equation}
  g_{WW'W'} = g_{WWW},
\label{eq:case1}  
\end{equation}
or
\begin{equation}
  g_{WW'W'} = -\xi_V g_{W'W'W'} \dfrac{M_W^2}{M_{W'}^2}.
\label{eq:case2}  
\end{equation}

We next turn to the properties of Higgs couplings 
$g_{WWh_m}$ and $g_{WW'h_m}$.
Let us start with the $E^2$ sum rule for the $WW \to WW$ amplitude.
Using (\ref{eq:e2-sr1}) we obtain
\begin{eqnarray}
  \sum_m g_{WWh_m}^2 &=& 4 M_W^2 (g_{WWW}^2 + g_{WWW'}^2)
                   \nonumber\\
                   & & - 3M_W^2 g_{WWW}^2 - 3M_{W'}^2 g_{WWW'}^2 ,
                   \nonumber\\
                   & &
\label{eq:wwh-sr}
\end{eqnarray}
where (\ref{eq:e4-sr1}) is also used to rewrite the quartic 
vector boson coupling strength $g_{WWWW}$ in terms of 
$g_{WWW}$ and $g_{WWW'}$.
Plugging (\ref{eq:wwwp-parametrization}) into (\ref{eq:wwh-sr}),
we find
\begin{equation}
  \sum_m g_{WWh_m}^2 = M_W^2 g_{WWW}^2 \left[
    1 - 3\xi_V^2 \dfrac{M_W^2}{M_{W'}^2}
  \right].
\label{eq:wwh-sr2}
\end{equation}
The Higgs boson coupling with the $WW$ state is therefore
affected by the parameter $\xi_V$ and the $W'$ boson mass.
The roles of the unitarity sum rule (\ref{eq:wwh-sr2})
have been widely studied in 
Refs.~\cite{Foadi:2008xj,Hernandez:2010iu,Bellazzini:2012tv,Englert:2015oga}.

The $E^2$ sum rules for the $WW \to WW'$ amplitude can also 
be derived from (\ref{eq:e2-sr1}),
\begin{eqnarray}
\lefteqn{
  \sum_m g_{WWh_m} g_{WW'h_m}
} \nonumber\\
   &=& (3M_W^2+M_{W'}^2) \left[ g_{WWW}g_{WW'W} + g_{WWW'} g_{WW'W'} \right]
   \nonumber\\
   & & - 3M_W^2 g_{WWW} g_{WW'W} - 3M_{W'}^2 g_{WWW'} g_{WW'W'}
   \nonumber\\
   &=& M_{W'}^2 g_{WWW} g_{WWW'} 
   \nonumber\\
   & & + (3M_W^2-2M_{W'}^2) g_{WWW'} g_{WW'W'}.
\label{eq:wwph-sr}
\end{eqnarray}
Putting the parametrization of $g_{WWW'}$ (\ref{eq:wwwp-parametrization}) and 
one solution of $g_{WW'W'}$ (\ref{eq:case1}) 
into (\ref{eq:wwph-sr}), we obtain
\begin{equation}
  \sum_m g_{WWh_m} g_{WW'h_m} = -\xi_V M_W^2 g_{WWW}^2 \left[
    1 + {\cal O}(\dfrac{M_W^2}{M_{W'}^2})
  \right].
\label{eq:wwph-sr-case1}
\end{equation}
If we put (\ref{eq:wwwp-parametrization}) and
the other solution of $g_{WW'W'}$ (\ref{eq:case2}), we find
\begin{equation}
  \sum_m g_{WWh_m} g_{WW'h_m} = \xi_V M_W^2 g_{WWW}^2 \left[
    1 + {\cal O}(\dfrac{M_W^2}{M_{W'}^2})
  \right].
\label{eq:wwph-sr-case2}
\end{equation}

For the $WW \to W'W'$ scattering, not only (\ref{eq:e2-sr1}) but also
(\ref{eq:e2-sr2}) gives non-trivial sum rules.
From (\ref{eq:e2-sr1}), we obtain
\begin{eqnarray}
\lefteqn{
  \sum_m g_{WW' h_m}^2 
} \nonumber\\
  &=& (2M_W^2 + 2M_{W'}^2) \left[ g_{WW'W}^2 + g_{WW'W'}^2\right]
  \nonumber\\
  & & - 2 \left[ M_W^2 g_{WW'W}^2 +M_{W'}^2 g_{WW'W}^2 \right]
  \nonumber\\
  & & - \left[ M_W^2 g_{WWW} g_{W'W'W} + M_{W'}^2 g_{WWW'} g_{W'W'W'} \right]
  \nonumber\\
  &=& 2M_{W'}^2 g_{WWW'}^2 + 2M_{W}^2 g_{WW'W'}^2
  \nonumber\\
  & & - \left[ M_W^2 g_{WWW} g_{WW'W'} + M_{W'}^2 g_{WWW'} g_{W'W'W'} 
        \right].
  \nonumber\\
  & &
\label{eq:wwph-wwph-sr}
\end{eqnarray}
We also obtain from (\ref{eq:e2-sr2})
\begin{eqnarray}
\lefteqn{
  \dfrac{(M_W^2-M_{W'}^2)^2}{M_W^2} g_{WW'W}^2
 +\dfrac{(M_W^2-M_{W'}^2)^2}{M_{W'}^2} g_{WW'W'}^2
} \nonumber\\
  &=& M_W^2 (g_{WW'W}^2 - g_{WWW} g_{W'W'W})
  \nonumber\\
  & & +M_{W'}^2 (g_{WW'W'}^2 - g_{WWW'} g_{W'W'W'}),
\end{eqnarray}
which reads
\begin{eqnarray}
\lefteqn{
M_W^2 g_{WWW} g_{WW'W'} + M_{W'}^2 g_{WWW'} g_{W'W'W'} 
} \nonumber\\
  &=& \left[ M_W^2 - \dfrac{(M_W^2-M_{W'}^2)^2}{M_W^2} \right] g_{WWW'}^2
  \nonumber\\
  & & +\left[ M_{W'}^2 - \dfrac{(M_W^2-M_{W'}^2)^2}{M_{W'}^2} \right] g_{WW'W'}^2 .
\label{eq:wwwpwp-sr2}
\end{eqnarray}
Note that the last line of (\ref{eq:wwph-wwph-sr}) can be erased
by using (\ref{eq:wwwpwp-sr2}).
We now have
\begin{equation}
  \label{eq:3}
  \sum_m g_{WW'h_m}^2 = \dfrac{M_{W'}^4}{M_W^2} g_{WWW'}^2
                     +\dfrac{M_{W}^4}{M_{W'}^2} g_{WW'W'}^2.
\end{equation}
It is now easy to show
\begin{equation}
  \sum_m g_{WW'h_m}^2 = \xi_V^2 M_W^2 g_{WWW}^2 \left[
    1 + {\cal O}(\dfrac{M_W^2}{M_{W'}^2})
  \right].
\label{eq:wwph2-sr}
\end{equation}

Combining (\ref{eq:wwh-sr2}), (\ref{eq:wwph-sr-case1}) 
and (\ref{eq:wwph2-sr}), we obtain
an impressive sum rule
\begin{equation}
  \dfrac{1}{M_W^2} \sum_m (g_{WW'h_m}+\xi_V g_{WWh_m})^2
  = \xi_V^2 g_{WWW}^2 {\cal O}(\dfrac{M_W^2}{M_{W'}^2}) .
\end{equation}
Similarly, using (\ref{eq:wwph-sr-case2}) instead
of (\ref{eq:wwph-sr-case1}), we find
\begin{equation}
  \dfrac{1}{M_W^2} \sum_m (g_{WW'h_m}-\xi_V g_{WWh_m})^2
  = \xi_V^2 g_{WWW}^2 {\cal O}(\dfrac{M_W^2}{M_{W'}^2}) .
\end{equation}
For both cases, we obtain
\begin{equation}
  g_{WW' h_m} = \pm \xi_V \left[ g_{WW h_m} \pm g_{WW h}^{\rm SM} 
                              {\cal O}(\dfrac{M_W}{M_{W'}})
                        \right] .
  \label{eq:main-result}
\end{equation}
We used here $g_{WW h}^{\rm SM} \simeq g_{WWW} M_W$.
The Higgs ($h$) coupling with $WW'$ is therefore related with
$g_{WWW'}$ and $g_{WWh}$ through the relation (\ref{eq:main-result})
in the large $M_{W'}$ limit.
If the coupling of the 125 GeV Higgs boson ($h$) 
$g_{WWh} \simeq g_{WWh}^{\rm SM}$, we find
the uncertainty in 
(\ref{eq:main-result}) is about 4 \% ($=M_{W}/M_{W'}$)
for $M_{W'}=2$ TeV and therefore is negligibly small.

We note (\ref{eq:main-result}) is a novel relation,
which has not been pointed out in earlier references.
The relation provides us further information for the $W'$ 
boson properties
beyond the widely studied $WW\to WW$ sum rule (\ref{eq:wwh-sr2}).
We emphasize that we provided the proof of (\ref{eq:main-result})
in models with arbitrary number of the neutral Higgs bosons.
Another remark is that the conclusion (\ref{eq:main-result})
is unchanged even if we consider models containing $W''$ or higher 
KK resonances.
The result (\ref{eq:main-result}) can therefore be applied to a large
class of perturbative models which include the $W'$ particle
and neutral Higgs bosons.

\section{Unitarity inspired parametrization for $W'$ and $Z'$}
\label{sec:parametrization}

In the previous section, we have shown that the coupling
strengths of the $W'$ boson can be controlled by the perturbative
unitarity requirements in the custodial $SU(2)$ symmetric model.
Especially, we found
\begin{equation}
  g_{WWW'} = \xi_V g_{WWW} \dfrac{M_W^2}{M_{W'}^2},
  \qquad
  \xi_V \lesssim 15 ,
\label{eq:parametrization1}
\end{equation}
and
\begin{equation}
  g_{WW' h_m} = \pm \xi_V g_{WWh_m}.
\end{equation}
In this section, we consider effects of custodial $SU(2)$ symmetry
violation $M_{Z}\ne M_{W}$, $M_{Z'} \ne M_{W'}$, requiring
the high energy $E^2$-behavior of the longitudinal $W$ and $Z$ boson 
scattering amplitudes are custodial $SU(2)$ symmetric.
This requirement is justified because the weak hypercharge coupling
(the origin of the custodial symmetry violation) 
does not affect the $E^2$-behavior in the amplitudes at the tree level.
We find (\ref{eq:parametrization1}) needs to be modified 
as\footnote{
The default value of the PYTHIA~\cite{Sjostrand:2007gs} implementation
of the extended gauge model~\cite{Altarelli:1989ff}
corresponds to $\xi_V = M_W^2/M_Z^2$.
}
\begin{equation}
  g_{WZW'} = \xi_V g_W \dfrac{M_W M_Z}{M_{W'}^2}.
  \label{eq:parametrization2a}
\end{equation}
\begin{equation}
  g_{WWZ'} = \xi_V g_W \dfrac{M_W^2}{M_{Z'}^2} R,
\label{eq:parametrization2b}
\end{equation}
with $R(M_{W'}/M_{Z'})$ being a function of $M_{W'}/M_{Z'}$ satisfying
$R(1)=1$.
Here $g_W$ stands for the gauge coupling strength of the $W$ boson.

The Higgs couplings with the $WW'$ and $ZZ'$ states can also be
parametrized by using the custodial $SU(2)$ symmetry.
We obtain
\begin{equation}
  g_{WW' h} = \xi_h g_W M_W, 
\label{eq:custodial-formulaa}
\end{equation}
\begin{equation}
  g_{ZZ'h} = \xi_h g_W M_Z R,
\label{eq:custodial-formulab}
\end{equation}
with $h$ being the 125 GeV Higgs boson.  
The requirement of the perturbative unitarity then leads to
\begin{equation}
  \xi_h = \pm \kappa_V  \xi_V,\qquad
  \kappa_V \equiv \dfrac{g_{WWh}}{g_W M_W}
\label{eq:main-result2}
\end{equation}
as we have shown in the previous section.
It is now straightforward to show the relation
(\ref{eq:equivalence2}).
The relation (\ref{eq:main-result2}) 
and therefore (\ref{eq:equivalence2}) should be regarded
as one of the most important unitarity relations obtained
in this paper.
We are now able to describe the physics of $W$, $W'$ and 
Higgs by using the two parameters ($\kappa_V$ and $\xi_V$),
instead of the three ($\kappa_V$, $\xi_V$ and $\xi_h$).

In order to study collider phenomenologies of $W'$ and $Z'$ 
particles, we need to specify the couplings of $W'$ and $Z'$
with quarks and leptons.
In this paper we adopt an ansatz in which $W'$ and $Z'$ couple
with weak current of quarks (leptons) with universal coupling strength
$\xi_q g_W$ ($\xi_\ell g_W$).
An example of model satisfying this ansatz will be introduced in 
Sec.~\ref{sec:threesite}.

\section{Fit to the ATLAS diboson anomaly}
\label{sec:fit}

We are now ready to search the parameter region of
$\xi_V$, $\xi_q$, and $\xi_\ell$ so as to explain 
the ATLAS 2 TeV diboson anomaly in $W'$ models.
As we discussed in Sec.~\ref{sec:introduction}, 
we use
\begin{equation}
  \sum_{X}\sigma(pp \to X)  B_X(VV) \simeq 10 \mbox{ fb},
\label{eq:reference}
\end{equation}
as a reference value to explain the ATLAS 2 TeV diboson anomaly.
Here $X$ is a narrow width new particle 
having 2 TeV mass decaying to the $VV$ states, with $V$ being 
a weak gauge boson $W$ or $Z$.
The ATLAS Collaboration reported excesses not only in the 
$WZ$ category, but also in the $WW$ and $ZZ$ categories.
It has being claimed that there exist significant size of event 
contamination among $WZ$, $WW$ and $ZZ$ categories in the 
$JJ$ events~\cite{Fukano:2015hga,Thamm:2015csa}, however.
We therefore evaluate
\begin{equation}
  \sum_{X=W', Z'}\sigma(pp \to X)  B_X(VV) 
\label{eq:what-we-want}
\end{equation}
for a degenerated $W'$ and $Z'$ model ($M_{W'}=M_{Z'}=2$ TeV),
and
\begin{equation}
  \sigma(pp \to W')  B_{W'}(WZ) 
\end{equation}
for a non-degenerated model ($M_{Z'}>M_{W'}=2$ TeV),
and simply compare the number with the reference value (\ref{eq:reference}).

In addition to $\sigma B$, we also need to explain the narrow width
of the new particle $X$, typically smaller than the bin size of the
experiment\cite{Aad:2015owa}, 100 GeV.

The model explaining the diboson anomaly need to be consistent 
with the existing limits on the $W'$ and $Z'$ bosons.
Both ATLAS and CMS experiments report upper limits on the 
production cross section of $W'$ in its leptonic decay channels\cite{ATLAS:2014wra,Khachatryan:2014tva}.
For 2 TeV $W'$ boson search in $pp$ collisions at $\sqrt{s}=8$ TeV, 
the limit is
\begin{equation}
  \sigma(pp \to W') B_{W'}(\ell\nu) \lesssim 0.4 \mbox{ fb}.
\label{eq:limit-ellnu}
\end{equation}
Limits on the $Z' \to e^+ e^-, \mu^+ \mu^-$ are reported in
Refs.\cite{Aad:2014cka,Khachatryan:2014fba}.  
For $M_{Z'}=2$ TeV, 
these references give a limit
\begin{equation}
  \sigma(pp \to Z') B_{Z'}(\ell^+ \ell^-) \lesssim 0.2\mbox{ fb}.
\label{eq:limit-ellell}
\end{equation}
The LHC limits on the resonant dijet production can also be used to 
constrain $W'$ models.
Using the limits presented in
Refs.\cite{Aad:2014aqa,Khachatryan:2015sja}, we see
\begin{equation}
  \sum_{X=W', Z'} \sigma(pp \to X) B_X(2j)
 \lesssim 100 \mbox{ fb},
\label{eq:limit-dijet}
\end{equation}
for a degenerated $M_{W'}=M_{Z'}=2$ TeV model, and
\begin{equation}
  \sigma(pp \to W') B_{W'}(2j) 
 \lesssim 100 \mbox{ fb},
\end{equation}
for a non-degenerated model with $M_{W'}=2$ TeV.

Finally, the model needs to satisfy the limit on the
$W' \to Wh$ and $Z' \to Zh$ decay modes.
Here $h$ stands for the 125 GeV Higgs particle.
The limit quoted in Ref.\cite{Khachatryan:2015bma}
is
\begin{equation}
  \sum_{X=W', Z'} \sigma(pp \to X) B_{X}(Vh)
 \lesssim 7 \mbox{ fb},
\label{eq:limit-higgs-WZ}
\end{equation}
for $M_{W'}=M_{Z'}=2$ TeV, and
\begin{equation}
  \sigma(pp \to W') B_{W'}(Wh) 
 \lesssim 7 \mbox{ fb},
\label{eq:limit-higgs-W}
\end{equation}
for a non-degenerated model with $M_{W'}=2$ TeV.

\begin{figure}[t]
  \centering
  \includegraphics[width=7cm]{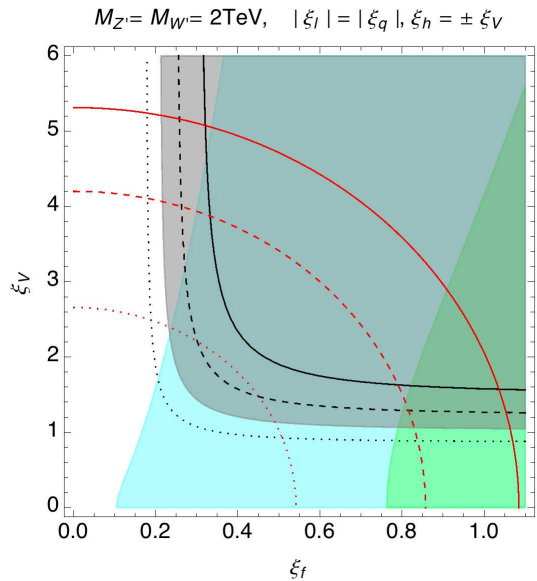}
  \caption{
Limits on the $W'$ and $Z'$ couplings in the $\xi_f$-$\xi_V$ plane
for the degenerated $M_{Z'}=M_{W'}=2$ TeV model.
$\xi_f=|\xi_q|=|\xi_\ell|$ and $\xi_h = \pm \xi_V$ are assumed.
The darkgreen region,
the lightblue region,
and the gray region
are excluded by the dijet mode limit (\ref{eq:limit-dijet}),
the $\ell \nu$ mode (\ref{eq:limit-ellnu}),
and Higgs mode (\ref{eq:limit-higgs-WZ}), respectively.
Although we do not show the limit from (\ref{eq:limit-ellell})
in the plot, it is numerically almost identical to the 
$W'\to \ell \nu$ limit.
The black solid curve, the black dashed curve, 
and the black dotted curve are for 
$\sigma(pp \to W') B_{W'}(WZ) + \sigma(pp \to Z') B_{Z'}(WW)=15$ fb, 
10 fb, and 5 fb, respectively.
The red solid curve, the red dashed curve,
and the red dotted curve are for $\Gamma_{W'}=80$GeV, 50GeV, and
20 GeV, respectively. The width of $Z'$ is almost equal to 
$\Gamma_{W'}$ thanks to the custodial symmetry.
}
  \label{fig:WZ-lquniv-h100}
\end{figure}

Figure~\ref{fig:WZ-lquniv-h100} shows these limits
in the $\xi_f$-$\xi_V$ plane for $M_{Z'}=M_{W'}=2$ TeV.
We assume quark-lepton universal coupling $\xi_f=|\xi_q|=|\xi_\ell|$
in this plot.
We also assume the 125 GeV Higgs coupling with $WW'$ and $ZZ'$
are given by
\begin{equation}
  \xi_h = \pm \xi_V ,
\label{eq:assumption1}
\end{equation}
which corresponds to $\kappa_V = 1$ in the 
unitarity relation (\ref{eq:main-result2}).
The dijet limit (\ref{eq:limit-dijet}) 
is applied for the resonant production cross section of 
five flavor $q\bar{q}$ pairs.
The $W'$ and $Z'$ particles are produced through their
couplings with quarks in $pp$ collisions at 8 TeV. 
The production cross sections are evaluated by using the 
CTEQ6L1 set of the parton distribution functions~\cite{Pumplin:2002vw}.

We see in this plot that 
the Higgs mode limit (\ref{eq:limit-higgs-WZ}) and
the leptonic decay mode limit (\ref{eq:limit-ellnu}) 
rule out huge parameter space.
It is impossible to obtain the reference value of the cross section
(\ref{eq:reference})
without causing conflicts with the Higgs mode limit (\ref{eq:limit-higgs-WZ})
under the Higgs coupling assumption (\ref{eq:assumption1}).
We are only able to achieve $\sum_{X=W', Z'} \sigma(X)B_X(VV)\simeq 7$ fb 
at most.

\begin{figure}[t]
  \centering
  \includegraphics[width=7cm]{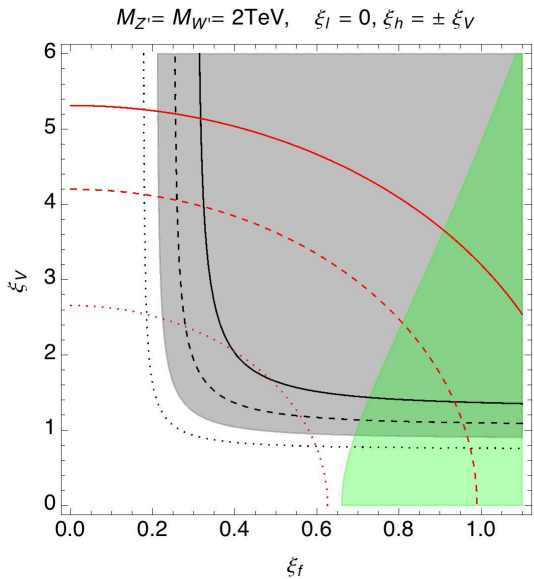}
  \caption{
Plot similar to Figure~\ref{fig:WZ-lquniv-h100}
for leptophobic case $\xi_\ell=0$, $\xi_f=|\xi_q|$.
$\xi_h = \pm \xi_V$ is assumed.
The darkgreen region,
and the gray region
are excluded by the dijet mode limit (\ref{eq:limit-dijet}),
and Higgs mode (\ref{eq:limit-higgs-WZ}), respectively.
The black solid curve, the black dashed curve, 
and the black dotted curve are for 
$\sigma(pp \to W') B_{W'}(WZ) + \sigma(pp \to Z') B_{Z'}(WW)=15$ fb, 
10 fb, and 5 fb, respectively.
The red solid curve, the red dashed curve,
and the red dotted curve are for $\Gamma_{W'}=80$GeV, 50GeV, and
20 GeV, respectively.
The width of $Z'$ is almost equal to 
$\Gamma_{W'}$ thanks to the custodial symmetry.
}
  \label{fig:WZ-lphobc-h100}
\end{figure}

A similar plot for a leptophobic $\xi_\ell=0$ model 
is shown in Figure~\ref{fig:WZ-lphobc-h100}
assuming the degenerated $W'$ and $Z'$, $M_{Z'}=M_{W'}=2$ TeV .
Again, the 125 GeV Higgs coupling is assumed to satisfy
(\ref{eq:assumption1}).
Although the constraints from the leptonic decay channels of $W'$ 
and $Z'$ disappear in the leptophobic model,
the limit on the $Vh$ channel gives severe constraint on 
(\ref{eq:what-we-want}).  It is impossible to obtain the
the reference value 10 fb in this setup.

\begin{figure}[t]
  \centering
  \includegraphics[width=7cm]{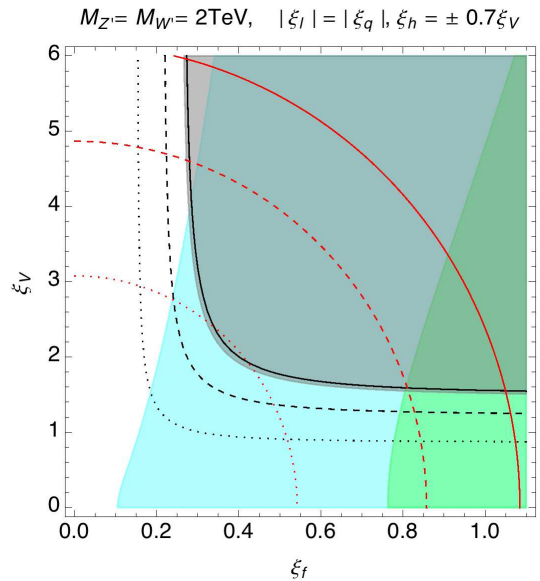}
  \caption{
Plot similar to Figure~\ref{fig:WZ-lquniv-h100} for
$\xi_h / \xi_V=\pm0.7$.
$\xi_f=|\xi_q|=|\xi_\ell|$ is assumed.
The darkgreen region,
the lightblue region,
and the gray region
are excluded by the dijet mode limit (\ref{eq:limit-dijet}),
the $\ell \nu$ mode (\ref{eq:limit-ellnu}),
and Higgs mode (\ref{eq:limit-higgs-WZ}), respectively.
Although we do not show the limit from (\ref{eq:limit-ellell})
in the plot, it is numerically almost identical to the 
$W'\to \ell \nu$ limit.
The black solid curve, the black dashed curve, 
and the black dotted curve are for 
$\sigma(pp \to W') B_{W'}(WZ) + \sigma(pp \to Z') B_{Z'}(WW)=15$ fb, 
10 fb, and 5 fb, respectively.
The red solid curve, the red dashed curve,
and the red dotted curve are for $\Gamma_{W'}=80$GeV, 50GeV, and
20 GeV, respectively.
The width of $Z'$ is almost equal to 
$\Gamma_{W'}$ thanks to the custodial symmetry.
}
  \label{fig:WZ-lquniv-h070}
\end{figure}

In order to explain the diboson excess without causing conflicts with
the Higgs mode limit (\ref{eq:limit-higgs-WZ}),
we need to take smaller value of $\xi_h$.
The unitarity relation (\ref{eq:main-result2}) suggests us such a 
value of $\xi_h$ can be achieved only if we consider models with
a non-SM like Higgs ($\kappa_V < 1$).
Figure~\ref{fig:WZ-lquniv-h070} shows the plot with $\xi_h/\xi_V=\pm0.7$, 
i.e., $\kappa_V=0.7$.
The quark-lepton universal couplings $\xi_f=|\xi_q|=|\xi_\ell|$ 
are assumed in the plot.
The reference cross section value for the
excess can be explained at, e.g.,  $\xi_V \simeq 4$
and $\xi_f \simeq 0.23$.
Note that the choice of this parameter $\xi_V$ satisfies the perturbativity
condition (\ref{eq:wwww-perturbativity2}).

\begin{figure}[t]
  \centering
  \includegraphics[width=7cm]{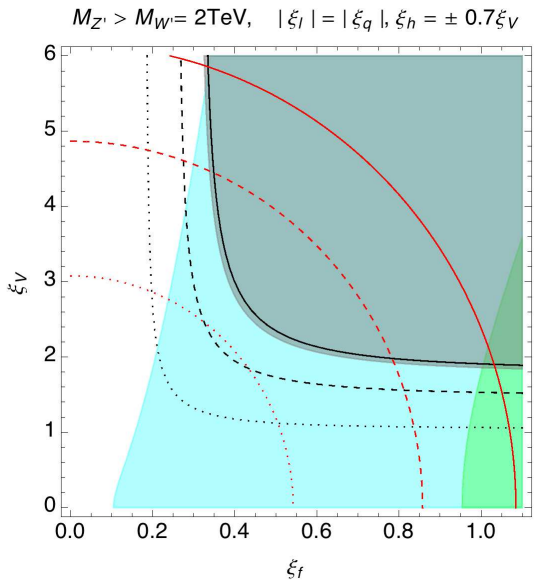}
  \caption{
Plot similar to Figure~\ref{fig:WZ-lquniv-h100} for
non-degenerated model $M_{Z'}>M_{W'}=2$TeV with $\xi_h / \xi_V=\pm0.7$.
$\xi_f=|\xi_q|=|\xi_\ell|$ is assumed.
The darkgreen region,
the lightblue region,
and the gray region
are excluded by the dijet mode limit (\ref{eq:limit-dijet}),
the $\ell \nu$ mode (\ref{eq:limit-ellnu}),
and Higgs mode (\ref{eq:limit-higgs-W}), respectively.
The black solid curve, the black dashed curve, 
and the black dotted curve are for 
$\sigma(pp \to W') B_{W'}(WZ)=15$ fb, 
10 fb, and 5 fb, respectively.
The red solid curve, the red dashed curve,
and the red dotted curve are for $\Gamma_{W'}=80$GeV, 50GeV, and
20 GeV, respectively.
}
  \label{fig:W-lquniv-h070}
\end{figure}

We next consider the non-degenerated case, $M_{Z'}>M_{W'}=2$ TeV.
The $Z'$ boson is assumed to be heavy enough to be separated from the
2 TeV resonance.
The plot corresponding to this model is shown in Figure~\ref{fig:W-lquniv-h070}.
Here $\xi_f=|\xi_\ell|=|\xi_q|$ and $\xi_h/\xi_V=\pm0.7$ are assumed.
We find that the
reference cross section value for the ATLAS diboson anomaly can be 
explained at, e.g.,
$\xi_V\simeq 4$ and $\xi_f \simeq 0.28$.

A plot similar to
Figures~\ref{fig:WZ-lquniv-h100}, \ref{fig:WZ-lphobc-h100}, 
and \ref{fig:WZ-lquniv-h070} 
is also presented in Ref.~\cite{Fukano:2015hga}
in the context of 
the techni-$\rho$ interpretation for the ATLAS diboson anomalies.
The plot presented in the latest version of Ref.~\cite{Fukano:2015hga}
seems to be consistent with our results.

\section{Three site moose model}
\label{sec:threesite}

So far, we have analyzed the unitarity sum rules and the 
interpretations of the LHC anomaly of the diboson resonance
in terms of $W'$ and $Z'$ without writing an explicit gauge 
invariant Lagrangian.

In this section, we introduce the three site moose 
model~\cite{Chivukula:2006cg}
as an example to explain the ATLAS diboson excess
in a perturbative manner.
We are able to check explicitly that the unitarity sum 
rules are satisfied in the three site moose model.
We also find that the parameter region explaining the
ATLAS diboson anomaly is naturally realized in this model.

The three site moose model\cite{Chivukula:2006cg}
 was originally introduced as
a deconstruction version of the Higgsless theory\cite{Csaki:2003dt}.
This model contains $W'$ and $Z'$ bosons as 
KK particles of electroweak gauge bosons.
These KK particles are, at least partly,
responsible for the unitarization of the longitudinal
weak gauge boson scattering amplitudes\cite{Csaki:2003dt,SekharChivukula:2001hz,Chivukula:2002ej}.
After the LHC discovery of the 125 GeV Higgs boson,
the three site moose model was extended to include 
the 125 GeV Higgs particle in Ref.~\cite{Abe:2012fb}.
This model can be regarded as a benchmark model to study
the phenomenologies of $W'$ and $Z'$ particles.

\begin{figure}[t]
  \centering
  \includegraphics[width=7cm]{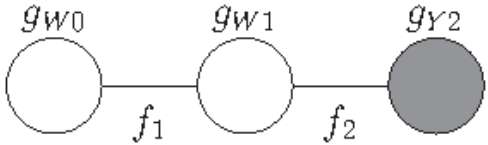}
  \caption{
The moose diagram for the three site model.
The blank circles represent $SU(2)$ gauge groups,
with coupling strengths $g_{W0}$ and $g_{W1}$, and
the shaded circle is a $U(1)$ gauge group with coupling
$g_{Y2}$.
}
  \label{fig:threesite-moose}
\end{figure}

The structure of gauge symmetry breaking in the three site moose model 
is illustrated in the ``moose notation''\cite{Georgi:1985hf} in
Figure~\ref{fig:threesite-moose}.  
In the three site moose model, 
we introduce $SU(2)_{W0} \times SU(2)_{W1} \times U(1)_{Y2}$ 
gauge groups.
The line connecting the $SU(2)_{W0}$ group with the $SU(2)_{W1}$ 
group in the moose diagram represents bi-fundamental ($2\times 2$ matrix)
Higgs field $\Phi_1$,
\begin{equation}
  \Phi_1 = S_1 \unit + i\tau^a \pi_1^a,
\label{eq:phi1}
\end{equation}
with $\tau^a$ being the Pauli spin matrix.
Similarly, the line between the $SU(2)_{W1}$ and $U(1)_{Y2}$ 
gauge groups corresponds to the Higgs field $\Phi_2$
\begin{equation}
  \Phi_2 = S_2 \unit + i\tau^a \pi_2^a .
\label{eq:phi2}
\end{equation}

The covariant derivatives of $\Phi_1$ and $\Phi_2$ are given 
by
\begin{equation}
  D_\mu \Phi_1 = \partial_\mu \Phi_1 + ig_{W0} W_{0\mu} \Phi_1
                                   - ig_{W1} \Phi_1 W_{1\mu},
\end{equation}
\begin{equation}
  D_\mu \Phi_2 = \partial_\mu \Phi_2 + ig_{W1} W_{1\mu} \Phi_2
                                   - ig_{Y2} \Phi_2 \frac{\tau^3}{2} B_{\mu}
  .
\end{equation}
Here $W_{0\mu} = W_{0\mu}^a \frac{\tau^a}{2}$, 
$W_{1\mu} = W_{1\mu}^a \frac{\tau^a}{2}$, 
and $B_\mu$ are the gauge fields of $SU(2)_{W0}$,
$SU(2)_{W1}$ and $U(1)_{Y2}$.

The Higgs field $\Phi_1$ is assumed to acquire its vacuum expectation
value (VEV),
\begin{equation}
  \label{eq:4}
  \VEV{\Phi_1} = f_1 \unit,
\end{equation}
which breaks the $SU(2)_{W0} \times SU(2)_{W1}$ into the diagonal 
subgroup $SU(2)$.
Similarly, the $SU(2)_{W1} \times U(1)_{Y2}$ is broken to $U(1)$
thanks to the VEV of $\Phi_2$,
\begin{equation}
  \VEV{\Phi_2} = f_2 \unit .
\end{equation}
Simultaneous existence of two VEVs $f_1\ne 0$ and $f_2\ne 0$ therefore
leads to the spontaneous symmetry breaking pattern 
\begin{equation}
  SU(2)_{W0} \times SU(2)_{W1} \times U(1)_{Y2} \to 
  U(1)_{\rm em} .
\end{equation}

Diagonalizing the mass matrices of $W_{0\mu}$, $W_{1\mu}$ and $B_\mu$
which arise from the Higgs kinetic term Lagrangian
\begin{equation}
  {\cal L}
  \ni \frac{1}{4} \tr \left[ (D_\mu \Phi_1)^\dagger (D^\mu \Phi_1) \right]
     +\frac{1}{4} \tr \left[ (D_\mu \Phi_2)^\dagger (D^\mu \Phi_2) \right],
\label{eq:higgs-kinetic}
\end{equation}
we obtain mass eigenstates, $W'$, $Z'$ (heavier massive charged and
neutral vector bosons), $W$, $Z$ (lighter massive vector bosons),
and a massless photon.

The weak hypercharge gauge couplings of quarks and leptons are given by
\begin{equation}
  {\cal L} \ni
  -J_{Y, {\rm quark}}^{\mu} B_\mu 
  -J_{Y, {\rm lepton}}^{\mu} B_\mu .
\end{equation} 
As shown in Ref.~\cite{Chivukula:2006cg}, after the gauge symmetry breaking,
the weak currents of quarks and leptons can be ``delocalized'' with
delocalization parameters $x_q$ and $x_\ell$ ($0 \le x_q \le 1$, 
$0 \le x_\ell \le 1$),
\begin{eqnarray}
  {\cal L} &\ni& -J_{W, {\rm quark}}^{a\mu} \left(
    g_{W0} W_{0\mu}^a (1-x_q) + g_{W1} W_{1\mu}^a x_q
  \right)
  \nonumber\\
  & & -J_{W, {\rm lepton}}^{a\mu} \left(
    g_{W0} W_{0\mu}^a (1-x_\ell)  +g_{W1} W_{1\mu}^a x_\ell 
  \right).
  \nonumber\\
  & &
\label{eq:weak-current}
\end{eqnarray}
We should emphasize here 
that the weak current of quarks and leptons couples
with both $SU(2)_{W0}$ and $SU(2)_{W1}$ in Eq.(\ref{eq:weak-current}).
This phenomenon called ``delocalization'' 
is characteristic in extra dimension scenarios\cite{Cacciapaglia:2004rb}, 
in which KK quarks and KK leptons exist.
Similar phenomenon also realizes in the partial 
compositeness scenarios~\cite{Kaplan:1991dc} in which the role
of the KK fermions is played by the composite fermions.
This is in contrast to the conventional G221 models 
 (including right-handed $SU(2)$ model), in which the weak current
cannot be delocalized. 

The delocalization parameters $x_q$ and $x_\ell$ need to be
flavor universal in order to avoid the flavor-changing-neutral-current
constraints in the three site model~\cite{Abe:2011sv}.
The quark delocalization parameter $x_q$ can differ from 
the lepton parameter $x_\ell$, however.
The electroweak precision constraints can be satisfied for heavy enough
$W'$ and $Z'$.
Even with lighter $M_{W'} \lesssim 1$ TeV,  
we are able to suppress the
Peskin-Takeuchi $S$-$T$ parameters\cite{Peskin:1990zt,Peskin:1991sw}
if we choose the lepton delocalization
parameter $x_\ell$ 
to the value determined by 
the ``ideal delocalization''~\cite{Chivukula:2005xm}.
As we will see later, the delocalization parameters
$x_q$ and $x_\ell$ are related with the $W'$ couplings with 
quarks and leptons $\xi_q$ and $\xi_\ell$.
Assuming the quark lepton universality of the $W'$ coupling
$\xi_q = \xi_\ell$, we are thus able to express the electroweak 
precision observable parameters in terms of 
$\xi_f = |\xi_q| = |\xi_\ell|$.
In this section, we will also check whether or not 
the region in the $\xi_f$-$\xi_V$ plane favored by the ATLAS 
diboson anomalies is consistent with the electroweak precision measurements.

We have two neutral Higgs bosons in this model.
One degree of freedom ($h_1$) of neural Higgs arises from $\Phi_1$,
the other ($h_2$) from $\Phi_2$.
The charged and pseudoscalar components of $\Phi_1$ and $\Phi_2$, i.e.,
$\pi_1^a$ and $\pi_2^a$ in (\ref{eq:phi1}) and (\ref{eq:phi2}) are all
eaten by massive gauge bosons $W$, $W'$, $Z$ and $Z'$.
The 125 GeV Higgs boson $h$ is considered to be a mixture of $h_1$ 
and $h_2$,
\begin{equation}
  h = h_1 \cos\alpha + h_2 \sin\alpha ,
\end{equation}
where
\begin{equation}
  S_1 = f_1 + h_1, \qquad
  S_2 = f_2 + h_2.
\end{equation}

We are now ready to discuss the ATLAS 2 TeV diboson anomaly in the 
three site moose model.
There are three possible ways to obtain the hierarchy 
$M_{Z'}, M_{W'}=2\mbox{ TeV} \gg M_{Z}, M_{W}$ in this setup.
One option is to take $g_{W1} \gg g_{W0}, g_{Y2}$ with keeping
the VEVs $f_1=f_2$ at the weak scale.
Collider phenomenologies in this option were studied\footnote{
Hadron collider phenomenologies of narrow spin-1 resonances 
in the technicolor models are studied 
in Refs.\cite{Eichten:1984eu,Lane:1991qh}.
The $f_1=f_2$ model is believed to be a low energy effective description
of the technicolor models.
See also Refs.\cite{Eichten:1996dx,Lane:1999uh,Lane:2002sm,Eichten:2007sx}
for collider phenomenologies of techni-vector mesons in the low scale 
technicolor model.}
in detail in Ref.~\cite{He:2007ge,Abe:2011qe,Du:2012vh}.
This limit is theoretically interesting, because it can be
regarded as an effective theory of strongly interacting Higgs 
sector~\cite{Casalbuoni:1985kq,Casalbuoni:1996qt,Lane:2009ct} motivated by
models of hidden local symmetry~\cite{Bando:1985ej,Bando:1985rf,Bando:1988ym,Bando:1988br,Harada:2003jx}. 

However, in order to realize 2 TeV $M_{W'}$ with this option,
we need non-perturbatively strong $g_{W1}$.
Ref.~\cite{Fukano:2015hga} studies an interpretation of the 
ATLAS diboson anomaly with $g_{W1} \gg g_{W0}, g_{Y2}$, introducing
higher order operators to suppress the effective coupling of the
heavy vector resonance.
We do not pursue this direction in this paper.

Other options are to take $f_1 \gg f_2$ or $f_1 \ll f_2$,
keeping perturbative coupling constants $g_{W0}$, $g_{W1}$
and $g_{Y2}$.
In the subsections below,  we will give our results 
of $M_{W'}$, $M_{Z'}$, $g_{WWZ'}$, $g_{WZW'}$, $g_{WWh}$
and $g_{WW'h}$ in these limits and
check the unitarity sum rules explicitly in this model.
We will also point out that 
the reciprocality between $\xi_f$ and $\xi_V$, suggested by
the favored parameter regions of the ATLAS diboson anomaly fit,
$\xi_V \simeq 3 \sim 5$, $\xi_f \simeq 0.2 \sim 0.3$ can be
naturally realized in this setup.

\subsection{$f_1 \gg f_2$}
We start with the case $f_1 \gg f_2$.
In this case the $SU(2)_{W0} \times SU(2)_{W1}$ gauge symmetry
is broken into the diagonal subgroup at the high energy scale $f_1$,
while the weak scale is given by $f_2$.
We thus obtain the masses of $W'$ and $Z'$ in proportional to
$f_1$,
\begin{equation}
  M_{W'}^2 \simeq M_{Z'}^2 \simeq \dfrac{g_{W0}^2 + g_{W1}^2}{4} f_1^2 .
\end{equation}
The weak $SU(2)$ gauge group at the weak scale 
should be the diagonal subgroup of $SU(2)_{W0}\times SU(2)_{W1}$, while 
the weak scale $U(1)_Y$ is given by $U(1)_{Y2}$.
The gauge coupling strengths at the weak scale are therefore 
\begin{equation}
  g_W^2 \simeq \dfrac{g_{W0}^2 g_{W1}^2}{g_{W0}^2+g_{W1}^2},
  \quad
  g_Y^2 \simeq g_{Y2}^2.
\end{equation}
The masses of the $W$ and $Z$ bosons are given by
\begin{equation}
  M_W^2 \simeq \dfrac{g_W^2}{4} f_2^2, \quad
  M_Z^2 \simeq \dfrac{g_W^2+g_Y^2}{4} f_2^2 .
\end{equation}
It is easy to check these formulas by an explicit diagonalization of the
mass matrices of the gauge fields $W_0$, $W_1$ and $B$,
which are given by 
the Higgs kinetic term Lagrangian (\ref{eq:higgs-kinetic}).
We also obtain
\begin{equation}
  M_{Z'}^2 - M_{W'}^2 = (M_Z^2-M_W^2) {\cal O}(\dfrac{M_W^2}{M_{W'}^2}).
\end{equation}
The $W'$ and $Z'$ bosons are therefore highly degenerated in this setup.
We next consider the $WZW'$ and $WWZ'$ couplings.
Explicit calculation of the mass diagonalization matrices of neutral
and charged gauge bosons leads to
\begin{equation}
  g_{WZW'} = \dfrac{g_{W1}}{g_{W0}} g_W \dfrac{M_W M_Z}{M_{W'}^2},
\label{eq:wzwp-caseA}
\end{equation}
and
\begin{equation}
  g_{WWZ'} = \dfrac{g_{W1}}{g_{W0}} g_W \dfrac{M_W^2}{M_{Z'}^2} .
\label{eq:wwzp-caseA}
\end{equation}
These results are perfectly consistent with our parametrization
formulas (\ref{eq:parametrization2a}) and (\ref{eq:parametrization2b}).
Therefore these couplings satisfy the unitarity and the custodial symmetry.
Comparing (\ref{eq:wzwp-caseA}), (\ref{eq:wwzp-caseA}) with 
(\ref{eq:parametrization2a}) and (\ref{eq:parametrization2b}), 
we find $\xi_V$ in this model is
given by
\begin{equation}
  \xi_V = \dfrac{g_{W1}}{g_{W0}} .
\end{equation}
We also obtain $R=1$, consistent with the custodial symmetry $M_{W'}=M_{Z'}$.

It is straightforward to calculate the Higgs couplings.
We obtain
\begin{equation}
  g_{WWh} \simeq g_W M_W \sin\alpha ,
\end{equation}
\begin{equation}
  g_{WW'h} \simeq -\dfrac{g_{W1}}{g_{W0}} g_W M_W \sin\alpha ,
\end{equation}
\begin{equation}
  g_{ZZ'h} \simeq -\dfrac{g_{W1}}{g_{W0}} g_W M_Z \sin\alpha .
\end{equation}
Again, these results are consistent with our
custodial symmetry formulas (\ref{eq:custodial-formulaa})
and  (\ref{eq:custodial-formulab})
and the result of the unitarity sum rules (\ref{eq:main-result2}).
The parameter $\xi_h$ and $\kappa_V$ in this model 
are given by
\begin{equation}
  \xi_h = -\dfrac{g_{W1}}{g_{W0}} \sin \alpha,
\end{equation}
\begin{equation}
  \kappa_V = \sin\alpha .
\end{equation}

We next calculate the $W'$ and $Z'$ couplings with
the quarks and the leptons.
We find that both the quark hypercharge current 
and the lepton hypercharge current
couple with the $Z'$ boson only with coefficients suppressed
by $(M_Z^2-M_W^2)/M_{W'}^2$.
The couplings of $W'$ and $Z'$ with the quarks and the 
leptons are therefore consistent with the ansatz given 
in Section.~\ref{sec:parametrization}.
The parameters $\xi_q$ and $\xi_\ell$ are given by
\begin{equation}
  \xi_q = \dfrac{g_{W0}}{g_{W1}} \left( 
    1- x_q - x_q \dfrac{g_{W1}^2}{g_{W0}^2}
  \right),
\end{equation}
\begin{equation}
  \xi_\ell = \dfrac{g_{W0}}{g_{W1}} \left( 
    1- x_\ell - x_\ell \dfrac{g_{W1}^2}{g_{W0}^2}
  \right),
\end{equation}
with $x_q$, $x_\ell$ being the delocalization parameters for
the quarks and the leptons.
We note that, if the delocalization parameters are small enough
$x_q \lesssim g_{W0}^2/g_{W1}^2$, 
$\xi_V$ and $\xi_q$ ($\xi_\ell$) satisfy the relation
\begin{equation}
  \xi_V \xi_q \lesssim 1 .
\end{equation}
Note that this relation is consistent with our result 
presented in Section.~\ref{sec:fit}.
The three site model with $g_{W0}/g_{W1}\simeq 0.25$ gives
the reference value cross section for the ATLAS diboson 
anomalies without causing conflicts 
with other limits on $W'$ and $Z'$.
The $\kappa_V$ is smaller than unity in this model, 
however. 
We need to take care the consistency with the signal strengths
in the Higgs production measurements as done in Ref.~\cite{Abe:2012fb}.

The electroweak precision observable parameters are evaluated at the tree
level as
\begin{equation}
  \hat{S} \simeq \dfrac{M_W^2}{M_{W'}^2} \xi_V \xi_f, \qquad
  W \simeq \dfrac{M_W^2}{M_{W'}^2} \xi_f^2.
\label{eq:Shat1}  
\end{equation}
Here we used the notation of Ref.\cite{Barbieri:2004qk}.
We assumed $\xi_f=|\xi_q|=|\xi_\ell|$ and neglected
terms suppressed by $1/\xi_V$ or $\xi_f$ in (\ref{eq:Shat1}),
given the situation that $\xi_V \simeq 3 \sim 4$ and 
$\xi_f \simeq 0.2 \sim 0.3$ are favored in the fit for the ATLAS 
diboson anomalies.
The analysis of Ref.\cite{Barbieri:2004qk} shows that 
$\hat{S}$ and $W$ need to be smaller than a few permil in order to
satisfy the electroweak precision constraints.
Note that both $\hat{S}$ and $W$ are suppressed by 
$M_W^2/M_{W'}^2 \simeq 1.6\times 10^{-3}$ in (\ref{eq:Shat1})
and vanish in the ideal delocalization limit $\xi_f=0$.
Thanks to the suppression factor $M_W^2/M_{W'}^2$, the explanation
of the ATLAS diboson anomaly is marginally consistent with the 
electroweak precision measurements at the tree level.

There also exist order of a few permil loop corrections to 
the electroweak precision parameters.
However, these loop corrections depends on the assumptions of
the UV completion behind the fermion delocalization.
For an example, the fermion delocalization can be UV-completed
by introducing additional heavy fermions.
The loop level corrections to the electroweak precision parameters
depends on the mass spectrum of the heavy additional fermions.
See, e.g., Ref.~\cite{Abe:2008hb}.
Since we do not specify such a UV-completion in this paper, we 
do not consider loop level constraints any further.

\subsection{$f_1 \ll f_2$}
The case $f_1 \ll f_2$ can be studied in a similar manner.
In this case the $SU(2)_{W1}\times U(1)_{Y2}$ group is broken 
to $U(1)_Y$ at the high energy scale $f_2$, while the
$SU(2)_{W0} \times U(1)_Y$ is broken by $f_1$ at the weak scale.
The masses of $W'$ and $Z'$ are therefore given by
\begin{equation}
  M_{W'}^2 \simeq \frac{1}{4} g_{W1}^2 f_2^2,
\end{equation}
\begin{equation}
  M_{Z'}^2 \simeq \frac{1}{4} (g_{W1}^2 +g_{Y2}^2) f_2^2 .
\end{equation}
The $Z'$ boson is heavier and can be separated from $W'$ at
the LHC experiments.
The model in this limit therefore corresponds to the non-degenerated case
in Section.~\ref{sec:fit} of this paper.
Since the weak hypercharge gauge boson at the weak scale is a
mixture of the $SU(2)_{W1}$ and $U(1)_{Y2}$ gauge bosons, 
the gauge coupling strengths at the weak scale are given by
\begin{equation}
  g_W^2 \simeq g_{W0}^2,
  \quad
  g_Y^2 \simeq \dfrac{g_{W1}^2 g_{Y2}^2}{g_{W1}^2+g_{Y2}^2},
\end{equation}
and we obtain the weak gauge boson 
mass,
\begin{equation}
  M_W^2 \simeq \dfrac{g_W^2}{4} f_1^2, \quad
  M_Z^2 \simeq \dfrac{g_W^2+g_Y^2}{4} f_1^2 .
\end{equation}

The $WZW'$, $WWZ'$, $WWh$, $WW'h$, $ZZ'h$ couplings are given by
\begin{equation}
  g_{WZW'} = \dfrac{g_{W1}}{g_{W0}} g_W \dfrac{M_W M_Z}{M_{W'}^2},
\end{equation}
\begin{equation}
  g_{WWZ'} = \dfrac{g_{W1}}{g_{W0}} g_W \dfrac{M_W^2}{M_{Z'}^2} 
           \dfrac{M_{W'}}{M_{Z'}} ,
\end{equation}
\begin{equation}
  g_{WWh} \simeq g_W M_W \cos\alpha,
\end{equation}
\begin{equation}
  g_{WW'h} \simeq \dfrac{g_{W1}}{g_{W0}} g_W M_W \cos\alpha ,
\end{equation}
\begin{equation}
  g_{ZZ'h} \simeq \dfrac{g_{W1}}{g_{W0}} g_W M_Z \dfrac{M_{W'}}{M_{Z'}} \cos\alpha ,
\end{equation}
which are consistent with our unitarity and custodial symmetry formulas
(\ref{eq:parametrization2a}), (\ref{eq:parametrization2b}), 
(\ref{eq:custodial-formulaa}), (\ref{eq:custodial-formulab}), and
(\ref{eq:main-result2}).
The parameters $\xi_V$, $\xi_h$,$\kappa_V$, and 
$R$ are given by
\begin{equation}
  \xi_V = \dfrac{g_{W1}}{g_{W0}}
\end{equation}
\begin{equation}
  \xi_h = \dfrac{g_{W1}}{g_{W0}} \cos \alpha
\end{equation}
\begin{equation}
  \kappa_V = \cos\alpha,
\end{equation}
and
\begin{equation}
  R = \dfrac{M_{W'}}{M_{Z'}}.
\end{equation}
The $Z'$ boson does couple with the fermion hypercharge currents
in this setup.
Therefore the couplings of $Z'$ with quarks and leptons cannot be
parametrized by the parameters $\xi_q$ and $\xi_\ell$.
The $Z'$ boson becomes heavier enough than $M_{W'}=2$ TeV, however,
and therefore irrelevant in the explanation of the 
ATLAS diboson anomalies.
In the phenomenological analysis, we therefore use the
parameters which describe the $W'$ coupling with quarks 
and leptons,
\begin{equation}
  \xi_q = - \dfrac{g_{W1}}{g_{W0}}  x_q ,
\end{equation}
\begin{equation}
  \xi_\ell = - \dfrac{g_{W1}}{g_{W0}}  x_\ell .
\end{equation}
We find that the relatively small value of $|\xi_q|\simeq 0.3$,
which is favored by the $W'$ constraints, 
is possible if we take $x_q \lesssim g_{W0}^2/g_{W1}^2$.

We find a non-decoupling tree level correction
\begin{equation}
  \hat{S} \simeq - \dfrac{\xi_f}{\xi_V},
\label{eq:Shat2}
\end{equation}
which is not suppressed by $M_W^2/M_{W'}^2$ in the $f_1 \ll f_2$ model
in contrast to the $f_1 \gg f_2$ case.
The constraint from the electroweak precision parameters 
is therefore much severer than the $f_1 \gg f_2$ case in this setup.
Again we assumed $\xi_f=|\xi_q|=|\xi_\ell|$ and neglected
terms suppressed by $1/\xi_V$ or $\xi_f$.
For $\xi_V \simeq 3 \sim 5$ and $\xi_f \simeq 0.2 \sim 0.3$, 
we find $\hat{S} \simeq 0.1$, clearly contradicting
with the present experimental limit on $\hat{S}$, namely 
$|\hat{S}| \lesssim 10^{-3}$~\cite{Barbieri:2004qk}.
The three site moose model with $f_1 \ll f_2$ is therefore ruled out
as an interpretation of the ATLAS diboson anomaly at least for the
quark-lepton universal coupling case $\xi_q=\xi_\ell$.

\section{Summary}
\label{sec:summary}

In this paper, we have studied general 
structures of perturbative $W'$ models from the
viewpoint of unitarity sum rules and the custodial
$SU(2)$ symmetry.
We found that the unitarity sum rules and the custodial 
symmetry are powerful enough, to predict many relations among
$WWZ'$, $WZW'$, $WWh$, $WW'h$ and $ZZ'h$ coupling strengths.
Especially, we derived a novel relation (\ref{eq:main-result}) 
from these sum rules, which 
can be applied to a large class of perturbative models 
including arbitrary numbers of the heavy vector triplet bosons 
and neutral Higgs bosons.
Using these relations, we surveyed parameter space of
$W'$ models to search for the region possible to
explain the ATLAS 2 TeV diboson anomalies.
If the CMS excesses at 1.8 TeV are not related with the
ATLAS 2 TeV anomalies, and if the 125 GeV Higgs boson is
the SM-like Higgs boson, we found that the CMS upper limit on 
the $W' \to Wh$ channel at 2 TeV is hardly compatible with 
the ATLAS 2 TeV diboson anomalies, suggesting non SM-like 
properties of the 125 GeV Higgs boson.
Based on the three site moose model Lagrangian, we then 
provided a couple of example models of non-SM like Higgs bosons, 
which may be able to explain the ATLAS diboson anomalies.

We emphasize that the three site moose model
we used  in this paper should be regarded merely 
as an example to illustrate the properties of
models which may explain the ATLAS 2 TeV diboson anomalies.
Models having similar properties (the non SM-like Higgs and 
the delocalization of the fermion weak current) such as 
the extra dimension models, the partial compositeness models\cite{Kaplan:1991dc},
 the top triangle moose models\cite{Chivukula:2009ck,Abe:2013jga}, and
the composite Higgs models\cite{Hong:2004td,Agashe:2004rs,Matsuzaki:2012gd,Lane:2014vca}
should be studied further.
Note that the proof of the relation (\ref{eq:main-result}) we provided 
in this paper is directly applicable only 
in a category of perturbative models with the custodial symmetry
containing arbitrary numbers of heavy vector triplet and 
neutral Higgs bosons.
Although this category of models already covers wide 
varieties of interesting models, there also exist 
phenomenologically viable  models which do not belong 
to this category.
It is particularly interesting to study the $W'$ phenomenology in these
non-perturbative models.

Each of ATLAS and CMS experiments will 
accumulate $\sim 10$ fb$^{-1}$ luminosity within year 2015 run 
at $\sqrt{s}=13$ TeV.
If the ATLAS diboson anomalies are settled to exist in these 
LHC Run2 experiments, the next target is to clarify the 
properties of the 2 TeV diboson resonance.
Especially, as we stressed in this paper, its $Vh$ decay channel
($V=W$ or $Z$) becomes important, since it can determine whether
the 125 GeV Higgs particle is SM-like or not.

\section*{Acknowledgements}
\label{sec:acknowledgements}

We thank Junji Hisano, 
Sekhar R. Chivukula,
Hidenori Fukano, Masafumi Kurachi, Shinya Matsuzaki 
and Koichi Yamawaki for useful discussions
and valuable comments. 
T.A.'s  work is supported by Grant-in-Aid for Scientific research from
the Ministry of Education, Culture, Sports, Science and Technology
(MEXT), Japan, No. 23104006.
R.N.'s work is supported by Research Fellowships of the Japan Society 
for the Promotion of Science (JSPS) for Young Scientists No.263947.
M.T.'s work is supported in part by the JSPS Grant-in-Aid 
for Scientific Research 15K05047.

\vspace*{2ex}
\section*{Note added}

After we sent our first version of manuscript to the arXiv,
a study on the unitarity implications for the ATLAS diboson 
anomalies had appeared~\cite{Cacciapaglia:2015eea},
in which the $WW \to WW$ sum rule (\ref{eq:wwh-sr2}) was used
to constrain the $WWh$ coupling.


\begin{thebibliography}{99}
\bibitem{Aad:2015owa}
  G.~Aad {\it et al.}  [ATLAS Collaboration],
  ``Search for high-mass diboson resonances with boson-tagged jets in proton-proton collisions at $\sqrt{s}$ = 8 TeV with the ATLAS detector,''
  arXiv:1506.00962 [hep-ex].

\bibitem{Khachatryan:2014hpa}
  V.~Khachatryan {\it et al.}  [CMS Collaboration],
  ``Search for massive resonances in dijet systems containing jets tagged as W or Z boson decays in pp collisions at $ \sqrt{s} $ = 8 TeV,''
  JHEP {\bf 1408} (2014) 173
  [arXiv:1405.1994 [hep-ex]].


\bibitem{Aad:2015ufa}
 G.~Aad {\it et al.}  [ATLAS Collaboration],
 ``Search for production of $WW/WZ$ resonances decaying to a lepton, neutrino 
  and jets in $pp$ collisions at $\sqrt{s}$ = 8 TeV with the ATLAS detector,''
 arXiv:1503.04677 [hep-ex].







\bibitem{Khachatryan:2015bma}
  V.~Khachatryan {\it et al.}  [CMS Collaboration],
  ``Search for a massive resonance decaying into a Higgs boson and a W or Z boson in hadronic final states in proton-proton collisions at sqrt(s) = 8 TeV,''
  arXiv:1506.01443 [hep-ex].

\bibitem{kn:CMS_higgs2}
 CMS Collaboration, 
 ``Search for massive $Wh$ resonances decaying to the $\ell\nu b\bar{b}$
   final state in the boosted regime at $\sqrt{s}=8$ TeV,''
 CMS PAS EXO-14-010.


\bibitem{Fukano:2015hga}
  H.~S.~Fukano, M.~Kurachi, S.~Matsuzaki, K.~Terashi and K.~Yamawaki,
  ``2 TeV Walking Technirho at LHC?,''
  arXiv:1506.03751 [hep-ph].

\bibitem{Hisano:2015gna}
  J.~Hisano, N.~Nagata and Y.~Omura,
  ``Interpretations of the ATLAS Diboson Resonances,''
  arXiv:1506.03931 [hep-ph].

\bibitem{Franzosi:2015zra}
  D.~B.~Franzosi, M.~T.~Frandsen and F.~Sannino,
  ``Diboson Signals via Fermi Scale Spin-One States,''
  arXiv:1506.04392 [hep-ph].

\bibitem{Cheung:2015nha}
  K.~Cheung, W.~Y.~Keung, P.~Y.~Tseng and T.~C.~Yuan,
  ``Interpretations of the ATLAS Diboson Anomaly,''
  arXiv:1506.06064 [hep-ph].


\bibitem{Dobrescu:2015qna}
  B.~A.~Dobrescu and Z.~Liu,
  ``A $W'$ boson near 2 TeV: predictions for Run 2 of the LHC,''
  arXiv:1506.06736 [hep-ph].

\bibitem{Aguilar-Saavedra:2015rna}
  J.~A.~Aguilar-Saavedra,
  ``Triboson interpretations of the ATLAS diboson excess,''
  arXiv:1506.06739 [hep-ph].

\bibitem{Alves:2015mua}
  A.~Alves, A.~Berlin, S.~Profumo and F.~S.~Queiroz,
  ``Dirac-Fermionic Dark Matter in $U(1)_X$ Models,''
  arXiv:1506.06767 [hep-ph].

\bibitem{Gao:2015irw}
  Y.~Gao, T.~Ghosh, K.~Sinha and J.~H.~Yu,
  ``G221 Interpretations of the Diboson and Wh Excesses,''
  arXiv:1506.07511 [hep-ph].

\bibitem{Thamm:2015csa}
  A.~Thamm, R.~Torre and A.~Wulzer,
  ``A composite Heavy Vector Triplet in the ATLAS di-boson excess,''
  arXiv:1506.08688 [hep-ph].

\bibitem{Brehmer:2015cia}
  J.~Brehmer, J.~Hewett, J.~Kopp, T.~Rizzo and J.~Tattersall,
  ``Symmetry Restored in Dibosons at the LHC?,''
  arXiv:1507.00013 [hep-ph].

\bibitem{Cao:2015lia}
  Q.~H.~Cao, B.~Yan and D.~M.~Zhang,
  ``Simple Non-Abelian Extensions and Diboson Excesses at the LHC,''
  arXiv:1507.00268 [hep-ph].

\bibitem{Cacciapaglia:2015eea}
  G.~Cacciapaglia and M.~T.~Frandsen,
  ``Unitarity implications of diboson resonance in the TeV region for Higgs physics,''
  arXiv:1507.00900 [hep-ph].


\bibitem{Pappadopulo:2014qza}
  D.~Pappadopulo, A.~Thamm, R.~Torre and A.~Wulzer,
  ``Heavy Vector Triplets: Bridging Theory and Data,''
  JHEP {\bf 1409} (2014) 060
  [arXiv:1402.4431 [hep-ph]].


\bibitem{Abe:2012fb}
  T.~Abe, N.~Chen and H.~J.~He,
  ``LHC Higgs Signatures from Extended Electroweak Gauge Symmetry,''
  JHEP {\bf 1301} (2013) 082
  [arXiv:1207.4103 [hep-ph]].

\bibitem{Chivukula:2006cg}
  R.~S.~Chivukula, B.~Coleppa, S.~Di Chiara, E.~H.~Simmons, H.~J.~He, M.~Kurachi and M.~Tanabashi,
  ``A Three Site Higgsless Model,''
  Phys.\ Rev.\ D {\bf 74} (2006) 075011
  [hep-ph/0607124].

\bibitem{ArkaniHamed:2001ca}
  N.~Arkani-Hamed, A.~G.~Cohen and H.~Georgi,
  ``(De)constructing dimensions,''
  Phys.\ Rev.\ Lett.\  {\bf 86} (2001) 4757
  [hep-th/0104005].

\bibitem{Hill:2000mu}
  C.~T.~Hill, S.~Pokorski and J.~Wang,
  ``Gauge invariant effective Lagrangian for Kaluza-Klein modes,''
  Phys.\ Rev.\ D {\bf 64} (2001) 105005
  [hep-th/0104035].

\bibitem{Csaki:2003dt}
  C.~Csaki, C.~Grojean, H.~Murayama, L.~Pilo and J.~Terning,
  ``Gauge theories on an interval: Unitarity without a Higgs,''
  Phys.\ Rev.\ D {\bf 69} (2004) 055006
  [hep-ph/0305237].


\bibitem{Cornwall:1973tb}
  J.~M.~Cornwall, D.~N.~Levin and G.~Tiktopoulos,
   ``Uniqueness of spontaneously broken gauge theories,''
  Phys.\ Rev.\ Lett.\  {\bf 30} (1973) 1268
   [Erratum-ibid.\  {\bf 31} (1973) 572].

\bibitem{Cornwall:1974km}
  J.~M.~Cornwall, D.~N.~Levin and G.~Tiktopoulos,
  ``Derivation of Gauge Invariance from High-Energy Unitarity Bounds on the s M
atrix,''
  Phys.\ Rev.\ D {\bf 10} (1974) 1145
   [Erratum-ibid.\ D {\bf 11} (1975) 972].

\bibitem{Llewellyn Smith:1973ey}
  C.~H.~Llewellyn Smith,
  ``High-Energy Behavior and Gauge Symmetry,''
  Phys.\ Lett.\ B {\bf 46} (1973) 233.

\bibitem{Gunion:1990kf}
  J.~F.~Gunion, H.~E.~Haber and J.~Wudka,
  ``Sum rules for Higgs bosons,''
  Phys.\ Rev.\ D {\bf 43} (1991) 904.

\bibitem{SekharChivukula:2008mj}
  R.~S.~Chivukula, H.~J.~He, M.~Kurachi, E.~H.~Simmons and M.~Tanabashi,
  ``General Sum Rules for WW Scattering in Higgsless Models: Equivalence Theorem and Deconstruction Identities,''
  Phys.\ Rev.\ D {\bf 78} (2008) 095003
  [arXiv:0808.1682 [hep-ph]].


\bibitem{Foadi:2008xj}
  R.~Foadi, M.~Jarvinen and F.~Sannino,
  ``Unitarity in Technicolor,''
  Phys.\ Rev.\ D {\bf 79} (2009) 035010
  [arXiv:0811.3719 [hep-ph]].

\bibitem{Hernandez:2010iu}
  A.~E.~Carcamo Hernandez and R.~Torre,
  ``A 'Composite' scalar-vector system at the LHC,''
  Nucl.\ Phys.\ B {\bf 841} (2010) 188
  [arXiv:1005.3809 [hep-ph]].

\bibitem{Bellazzini:2012tv}
  B.~Bellazzini, C.~Csaki, J.~Hubisz, J.~Serra and J.~Terning,
  ``Composite Higgs Sketch,''
  JHEP {\bf 1211} (2012) 003
  [arXiv:1205.4032 [hep-ph]].

\bibitem{Englert:2015oga}
  C.~Englert, P.~Harris, M.~Spannowsky and M.~Takeuchi,
  ``Unitarity-controlled resonances after Higgs discovery,''
  arXiv:1503.07459 [hep-ph].


\bibitem{Lee:1977eg}
  B.~W.~Lee, C.~Quigg and H.~B.~Thacker,
  ``Weak Interactions at Very High-Energies: The Role of the Higgs Boson Mass,'
'
  Phys.\ Rev.\ D {\bf 16} (1977) 1519.





\bibitem{Sjostrand:2007gs}
  T.~Sjostrand, S.~Mrenna and P.~Z.~Skands,
  ``A Brief Introduction to PYTHIA 8.1,''
  Comput.\ Phys.\ Commun.\  {\bf 178} (2008) 852
  [arXiv:0710.3820 [hep-ph]].


\bibitem{Altarelli:1989ff}
  G.~Altarelli, B.~Mele and M.~Ruiz-Altaba,
  ``Searching for New Heavy Vector Bosons in $p \bar{p}$ Colliders,''
  Z.\ Phys.\ C {\bf 45} (1989) 109
   [Z.\ Phys.\ C {\bf 47} (1990) 676].


\bibitem{ATLAS:2014wra}
  G.~Aad {\it et al.}  [ATLAS Collaboration],
  ``Search for new particles in events with one lepton and missing transverse momentum in $pp$ collisions at $\sqrt{s}$ = 8 TeV with the ATLAS detector,''
  JHEP {\bf 1409} (2014) 037
  [arXiv:1407.7494 [hep-ex]].

\bibitem{Khachatryan:2014tva}
  V.~Khachatryan {\it et al.}  [CMS Collaboration],
  ``Search for physics beyond the standard model in final states with a lepton and missing transverse energy in proton-proton collisions at sqrt(s) = 8 TeV,''
  Phys.\ Rev.\ D {\bf 91} (2015) 9,  092005
  [arXiv:1408.2745 [hep-ex]].

\bibitem{Aad:2014cka}
  G.~Aad {\it et al.}  [ATLAS Collaboration],
  ``Search for high-mass dilepton resonances in pp collisions at $\sqrt{s}=8$TeV with the ATLAS detector,''
  Phys.\ Rev.\ D {\bf 90} (2014) 5,  052005
  [arXiv:1405.4123 [hep-ex]].


\bibitem{Khachatryan:2014fba}
  V.~Khachatryan {\it et al.}  [CMS Collaboration],
  ``Search for physics beyond the standard model in dilepton mass spectra in proton-proton collisions at $ \sqrt{s}=8 $ TeV,''
  JHEP {\bf 1504} (2015) 025
  [arXiv:1412.6302 [hep-ex]].

\bibitem{Aad:2014aqa}
  G.~Aad {\it et al.}  [ATLAS Collaboration],
  ``Search for new phenomena in the dijet mass distribution using $p$-$p$ collision data at $\sqrt{s}=8$ TeV with the ATLAS detector,''
  Phys.\ Rev.\ D {\bf 91} (2015) 5,  052007
  [arXiv:1407.1376 [hep-ex]].

\bibitem{Khachatryan:2015sja}
  V.~Khachatryan {\it et al.}  [CMS Collaboration],
  ``Search for resonances and quantum black holes using dijet mass spectra in proton-proton collisions at $\sqrt{s} =$ 8 TeV,''
  Phys.\ Rev.\ D {\bf 91} (2015) 5,  052009
  [arXiv:1501.04198 [hep-ex]].


\bibitem{Pumplin:2002vw}
  J.~Pumplin, D.~R.~Stump, J.~Huston, H.~L.~Lai, P.~M.~Nadolsky and W.~K.~Tung,
  ``New generation of parton distributions with uncertainties from global QCD analysis,''
  JHEP {\bf 0207} (2002) 012
  [hep-ph/0201195].

\bibitem{SekharChivukula:2001hz}
  R.~S.~Chivukula, D.~A.~Dicus and H.~J.~He,
  ``Unitarity of compactified five-dimensional Yang-Mills theory,''
  Phys.\ Lett.\ B {\bf 525} (2002) 175
  [hep-ph/0111016].

\bibitem{Chivukula:2002ej}
  R.~S.~Chivukula and H.~J.~He,
  ``Unitarity of deconstructed five-dimensional Yang-Mills theory,''
  Phys.\ Lett.\ B {\bf 532} (2002) 121
  [hep-ph/0201164].

\bibitem{Georgi:1985hf}
H.~Georgi, ``A tool kit for builders of composite models,''  {Nucl.
  Phys.} {\bf B266} (1986) 274.

\bibitem{Cacciapaglia:2004rb}
  G.~Cacciapaglia, C.~Csaki, C.~Grojean and J.~Terning,
  ``Curing the Ills of Higgsless models: The S parameter and unitarity,''
  Phys.\ Rev.\ D {\bf 71} (2005) 035015
  [hep-ph/0409126].

\bibitem{Kaplan:1991dc}
  D.~B.~Kaplan,
  ``Flavor at SSC energies: A New mechanism for dynamically generated fermion masses,''
  Nucl.\ Phys.\ B {\bf 365} (1991) 259.

\bibitem{Abe:2011sv}
  T.~Abe, R.~S.~Chivukula, E.~H.~Simmons and M.~Tanabashi,
  ``The Flavor Structure of the Three-Site Higgsless Model,''
  Phys.\ Rev.\ D {\bf 85} (2012) 035015
  [arXiv:1109.5856 [hep-ph]].

\bibitem{Peskin:1990zt}
  M.~E.~Peskin and T.~Takeuchi,
  ``A New constraint on a strongly interacting Higgs sector,''
  Phys.\ Rev.\ Lett.\  {\bf 65} (1990) 964.

\bibitem{Peskin:1991sw}
  M.~E.~Peskin and T.~Takeuchi,
  ``Estimation of oblique electroweak corrections,''
  Phys.\ Rev.\ D {\bf 46} (1992) 381.


\bibitem{Chivukula:2005xm}
  R.~S.~Chivukula, E.~H.~Simmons, H.~J.~He, M.~Kurachi and M.~Tanabashi,
  ``Ideal fermion delocalization in Higgsless models,''
  Phys.\ Rev.\ D {\bf 72} (2005) 015008
  [hep-ph/0504114].


\bibitem{Eichten:1984eu}
  E.~Eichten, I.~Hinchliffe, K.~D.~Lane and C.~Quigg,
  ``Super Collider Physics,''
  Rev.\ Mod.\ Phys.\  {\bf 56} (1984) 579
   [Rev.\ Mod.\ Phys.\  {\bf 58} (1986) 1065].

\bibitem{Lane:1991qh}
  K.~D.~Lane and M.~V.~Ramana,
  ``Walking technicolor signatures at hadron colliders,''
  Phys.\ Rev.\ D {\bf 44} (1991) 2678.

\bibitem{Eichten:1996dx}
  E.~Eichten and K.~D.~Lane,
  ``Low - scale technicolor at the Tevatron,''
  Phys.\ Lett.\ B {\bf 388} (1996) 803
  [hep-ph/9607213].


\bibitem{Lane:1999uh}
  K.~D.~Lane,
  ``Technihadron production and decay in low scale technicolor,''
  Phys.\ Rev.\ D {\bf 60} (1999) 075007
  [hep-ph/9903369].

\bibitem{Lane:2002sm}
  K.~Lane and S.~Mrenna,
  ``The Collider phenomenology of technihadrons in the technicolor straw man model,''
  Phys.\ Rev.\ D {\bf 67} (2003) 115011
  [hep-ph/0210299].

\bibitem{Eichten:2007sx}
  E.~Eichten and K.~Lane,
  ``Low-scale technicolor at the Tevatron and LHC,''
  Phys.\ Lett.\ B {\bf 669} (2008) 235
  [arXiv:0706.2339 [hep-ph]].


\bibitem{He:2007ge}
  H.~J.~He, Y.~P.~Kuang, Y.~H.~Qi, B.~Zhang, A.~Belyaev, R.~S.~Chivukula, N.~D.~Christensen and A.~Pukhov {\it et al.},
  ``CERN LHC Signatures of New Gauge Bosons in Minimal Higgsless Model,''
  Phys.\ Rev.\ D {\bf 78} (2008) 031701
  [arXiv:0708.2588 [hep-ph]].

\bibitem{Abe:2011qe}
  T.~Abe, T.~Masubuchi, S.~Asai and J.~Tanaka,
  ``Drell-Yan Production of Z' in the Three-Site Higgsless Model at the LHC,''
  Phys.\ Rev.\ D {\bf 84} (2011) 055005
  [arXiv:1103.3579 [hep-ph]].


\bibitem{Du:2012vh}
  C.~Du, H.~J.~He, Y.~P.~Kuang, B.~Zhang, N.~D.~Christensen, R.~S.~Chivukula and E.~H.~Simmons,
  ``Discovering New Gauge Bosons of Electroweak Symmetry Breaking at LHC-8,''
  Phys.\ Rev.\ D {\bf 86} (2012) 095011
  [arXiv:1206.6022 [hep-ph]].

\bibitem{Casalbuoni:1985kq}
R.~Casalbuoni, S.~De~Curtis, D.~Dominici, and R.~Gatto, ``Effective weak
  interaction theory with possible new vector resonance from a strong higgs
  sector,''  {Phys. Lett.} {\bf B155} (1985) 95.

\bibitem{Casalbuoni:1996qt}
R.~Casalbuoni {\em et.~al.}, ``Degenerate bess model: The possibility of a
  low energy strong electroweak sector,''  {Phys. Rev.} {\bf D53} (1996)
  5201--5221, [\href{http://xxx.lanl.gov/abs/hep-ph/9510431}{{\tt
  hep-ph/9510431}}].


\bibitem{Lane:2009ct}
  K.~Lane and A.~Martin,
  ``An Effective Lagrangian for Low-Scale Technicolor,''
  Phys.\ Rev.\ D {\bf 80} (2009) 115001
  [arXiv:0907.3737 [hep-ph]].


\bibitem{Bando:1985ej}
M.~Bando, T.~Kugo, S.~Uehara, K.~Yamawaki, and T.~Yanagida, ``Is rho meson a
  dynamical gauge boson of hidden local symmetry?,''  {Phys. Rev. Lett.}
  {\bf 54} (1985) 1215.

\bibitem{Bando:1985rf}
M.~Bando, T.~Kugo, and K.~Yamawaki, ``On the vector mesons as dynamical
  gauge bosons of hidden local symmetries,''  {Nucl. Phys.} {\bf B259}
  (1985) 493.


\bibitem{Bando:1988ym}
M.~Bando, T.~Fujiwara, and K.~Yamawaki, ``Generalized hidden local symmetry
  and the a1 meson,''  {Prog. Theor. Phys.} {\bf 79} (1988) 1140.

\bibitem{Bando:1988br}
M.~Bando, T.~Kugo, and K.~Yamawaki, ``Nonlinear realization and hidden local
  symmetries,''  {Phys. Rept.} {\bf 164} (1988) 217--314.

\bibitem{Harada:2003jx}
M.~Harada and K.~Yamawaki, ``Hidden local symmetry at loop: A new
  perspective of composite gauge boson and chiral phase transition,''  {
  Phys. Rept.} {\bf 381} (2003) 1--233,
  [{\tt hep-ph/0302103}].


\bibitem{Barbieri:2004qk}
  R.~Barbieri, A.~Pomarol, R.~Rattazzi and A.~Strumia,
  ``Electroweak symmetry breaking after LEP-1 and LEP-2,''
  Nucl.\ Phys.\ B {\bf 703} (2004) 127
  [hep-ph/0405040].

\bibitem{Abe:2008hb}
  T.~Abe, S.~Matsuzaki and M.~Tanabashi,
  ``Does the three site Higgsless model survive the electroweak precision tests at loop?,''
  Phys.\ Rev.\ D {\bf 78} (2008) 055020
  [arXiv:0807.2298 [hep-ph]].


\bibitem{Chivukula:2009ck}
  R.~Sekhar Chivukula, N.~D.~Christensen, B.~Coleppa and E.~H.~Simmons,
  ``The Top Triangle Moose: Combining Higgsless and Topcolor Mechanisms for Mass Generation,''
  Phys.\ Rev.\ D {\bf 80} (2009) 035011
  [arXiv:0906.5567 [hep-ph]].

\bibitem{Abe:2013jga}
  T.~Abe and R.~Kitano,
  ``Phenomenology of Partially Composite Standard Model,''
  Phys.\ Rev.\ D {\bf 88} (2013) 1,  015019
  [arXiv:1305.2047 [hep-ph]].

\bibitem{Hong:2004td}
  D.~K.~Hong, S.~D.~H.~Hsu and F.~Sannino,
  ``Composite Higgs from higher representations,''
  Phys.\ Lett.\ B {\bf 597} (2004) 89
  [hep-ph/0406200].

\bibitem{Agashe:2004rs}
  K.~Agashe, R.~Contino and A.~Pomarol,
  ``The Minimal composite Higgs model,''
  Nucl.\ Phys.\ B {\bf 719} (2005) 165
  [hep-ph/0412089].

\bibitem{Matsuzaki:2012gd}
  S.~Matsuzaki and K.~Yamawaki,
  ``Techni-dilaton at 125 GeV,''
  Phys.\ Rev.\ D {\bf 85} (2012) 095020
  [arXiv:1201.4722 [hep-ph]].


\bibitem{Lane:2014vca}
  K.~Lane,
  ``A composite Higgs model with minimal fine-tuning: The large-$N$ and weak-technicolor limit,''
  Phys.\ Rev.\ D {\bf 90} (2014) 9,  095025
  [arXiv:1407.2270 [hep-ph]].

\end{thebibliography}
\end{document}